\documentclass[12pt,tightenlines,eqsecnum,floats,aps,amsmath,amssymb,nofootinbib,prd,floatfix]{revtex4}

\usepackage{setspace}
\usepackage{subfigure}
\usepackage{amsmath,amssymb,amsfonts,amsthm,mathrsfs}
\usepackage{graphicx,wrapfig}
\usepackage{enumerate}
\usepackage{color,xcolor}

\begin{document}

\title{Spherically symmetric loop quantum gravity: analysis of improved dynamics}

\author{Rodolfo Gambini$^{1}$, Javier Olmedo$^{2,3}$, Jorge Pullin$^{2}$}
\affiliation{
  1. Instituto de Física, Facultad de Ciencias, Universidad de la República, Montevideo, 11400, Uruguay \\
  2. Department of Physics and Astronomy, Louisiana State University,
  Baton Rouge, LA 70803-4001\\
  3. Departamento de F\'isica Te\'orica y del Cosmos, Universidad de Granada,  Granada-18071, Spain}

\begin{abstract}
We study the ``improved dynamics" for the treatment of spherically symmetric space-times in loop quantum gravity introduced by Chiou {\em et al.} in analogy with the one that has been constructed by Ashtekar, Pawlowski and Singh for the homogeneous space-times. In this dynamics the polymerization parameter is a well motivated function of the dynamical variables, reflecting the fact that the quantum of area depends on them. Contrary to the homogeneous case, its implementation does not trigger undesirable physical properties. We identify semiclassical physical states in the quantum theory and derive the corresponding effective semiclassical metrics. We then discuss some of their properties. Concretely, the space-time approaches sufficiently fast the Schwarzschild geometry at low curvatures. Besides, regions where the singularity is in the classical theory get replaced by a regular but discrete effective geometry with finite and Planck order curvature, regardless of the mass of the black hole. This circumvents trans-Planckian curvatures that appeared for astrophysical black holes in the quantization scheme without the improvement. It makes the resolution of the singularity more in line with the one observed in models that use the isometry of the interior of a Schwarzschild black hole with the Kantowski--Sachs loop quantum cosmologies. One can observe the emergence of effective violations of the null energy condition
in the interior of the black hole as part of the mechanism of the elimination of the singularity.
\end{abstract}
\maketitle
\section{Introduction}

Over the last two decades, loop quantum gravity inspired quantization techniques have been applied to situations with high degree of symmetry. Initial investigations concentrated on homogeneous cosmologies (see \cite{ashtekarsingh} for a review), but progress has also been made in the case of spherically symmetric space-times (see \cite{us,shell,us-rn,bh-rev}), space-times with two Killing vector fields \cite{madrid,uskerr} and even dilatonic black holes \cite{cghs}. Considerable parallel efforts to derive effective models from the full theory are on going \cite{qrlg,qrlg-bh,dap-lie,adl,path-int}.  In all these situations an important element in the quantization is the procedure known as ``polymerization". In it, some variables get replaced by exponentiated versions of them including a parameter. In the limit in which the parameter goes to zero one recovers the classical expressions. This procedure is inspired in the nature of the loop variables used in loop quantum gravity. For instance, one can approximate the curvature of a connection by considering the holonomy along a loop that shrinks to a point. However, in full loop quantum gravity we know that areas are quantized: there is a minimum quantum of area. Therefore one cannot take the limit in which an infinitesimal loop shrinks to a point. More precisely, in the Hilbert space of loop quantum gravity holonomies are well defined but connections and curvatures are not. Inspired by this observation, one proceeds in a similar way in the symmetry reduced models. One keeps the parameter in the variables that one polymerizes finite. In initial explorations such parameter was taken to be a constant. Later however, it was observed that the value of the quantum of area is dependent on the canonical variables. In the context of homogeneous cosmologies this had important implications. Quantizations with the fixed polymerization parameter (known as ``$\mu_0$" schemes) had a significant drawback: one could make non-trivial departures from classical general relativity occur at arbitrarily low curvatures. This was clearly undesirable, as one does not expect quantum gravity effects to appear at low curvatures. It was observed that if one made the polymerization parameter dependent on the canonical variables (known as $\bar{\mu}$ scheme) such a pathology was corrected. Other pathologies as the dependence of results on the ``fiducial cell'' in some models were also resolved. This ``improved quantization" scheme \cite{aspasi} has become the standard of quantization in the context of cosmology ever since.

Up to present, most investigations done directly in spherical symmetry (there were other investigations that exploit the isometry of the interior of Schwarzschild to Kantowski--Sachs, see \cite{aos} and references therein) were done with a fixed polymerization parameter. In spite of that, there was no analogue of the pathologies that emerged in the homogeneous case. The spherically symmetric space-times see departures from the behavior in classical general relativity in regions where curvatures are very large (close to the singularity), but this departure is suitably tamed. However, the values of curvature that were achieved were, for macroscopic black holes, trans-Planckian. This is not in line with what happens in loop quantum cosmology, and therefore with treatments of the Schwarzschild interior based on the isometry with Kantowski--Sachs.

It is therefore of interest to explore what implications there are in spherically symmetric space-times if one uses the generalization to that context the ``$\bar{\mu}$" style quantization. In the context of inhomogeneous spherically symmetric space-times these ideas have been discussed in some detail in Ref. \cite{chiou} from the point of view of an effective description (and motivated by a detailed kinematical analysis). However, a detailed implementation in the quantum theory of Ref. \cite{us} and a detailed derivation of effective semiclassical geometries have not been studied. This will be the purpose of this paper. We will construct a suitable generalization of the technique and apply it to spherically symmetric space-times. We will see that the technical aspects of the resulting quantization differ very little from the one constructed with a fixed polymerization parameter. Physically, on the other hand, the main effect is to limit the upper bounds in the curvature to Planck scale in the region that replaces the classical singularity. In the $\mu_0$ style quantization much larger trans-Planckian values were present in macroscopic black hole cases. In that sense the new scheme aligns better with what has been observed in singularity elimination in loop quantum cosmology and in spherically symmetric treatments that exploit the isometry between the Schwarzschild interior and Kantowski--Sachs loop quantum cosmologies. We also explore these and other geometrical properties of the resulting space-time by constructing an effective classical metric out of suitable semiclassical physical quantum states. Moreover, we characterize the departure of the effective metric from classical GR in an effective stress-energy tensor and discuss its properties. One very naturally sees the emergence of effective negative local mass and violations of the energy conditions as part of the elimination of the singularity.

The paper is organized as follows. In Sec. \ref{sec:class} we introduce the classical setting. The kinematical aspects of the quantum theory and the improved dynamics scheme is described in Sec. \ref{sec:kin-impd}. Sec. \ref{sec:phys} is devoted to the physical Hilbert space and observables. The main properties of the effective metric are discussed in Sec. \ref{sec:efft}. We conclude in Sec. \ref{sec:concl}. Besides, we added Appendices \ref{app:class-EF}, \ref{app2} and  \ref{app:cont} for the sake of completeness.

\section{Classical theory}\label{sec:class}

The classical theory (see \cite{us} for details) consists of a spherically symmetric spacetime metric given by
\begin{equation}
ds^2=-(N^2-N_xN^x)dt^2+2N_xdtdx+\frac{(E^\varphi)^2}{|E^x|}dx^2+|E^x| d\omega^2,
\end{equation}
where $d\omega^2$ is the line element of the unit 2-sphere, $N$ and $N_x$ are suitable lapse and shift functions, with $N^x=g^{xx}N_x$, and $E^\varphi$ and $E^x$ triad variables, conjugated to $K_\varphi$ and $K_x$, with Poisson brackets
\begin{align}\nonumber\label{eq:poiss}
&\{K_x(x),E^x(x')\}=G\delta(x-x'),\\
&\{{K}_\varphi(x),E^\varphi(x')\}=G\delta(x-x').
\end{align} 
We take the Immirzi parameter $\gamma=1$. The dynamics of these phase space variables are subject to the total Hamiltonian
\begin{equation}\label{eq:total-ham}
H_T=\int dx (NH+N^xH_x),
\end{equation}
with 
\begin{align}
	& H_x:=G^{-1}[E^\varphi K_\varphi'-(E^x)' K_x]\,,\label{eq:difeo}\\ \nonumber
	&H :=G^{-1}\left\{\frac{\left[(E^x)'\right]^2}{8\sqrt{E^x}E^\varphi}
	-\frac{E^\varphi}{2\sqrt{E^x}} - 2 K_\varphi \sqrt{E^x} K_x  
	-\frac{E^\varphi K_\varphi^2}{2 \sqrt{E^x}}\right.\\
	&\left.-\frac{\sqrt{E^x}(E^x)' (E^\varphi)'}{2 (E^\varphi)^2} +
	\frac{\sqrt{E^x} (E^x)''}{2 E^\varphi}\right\}\,,\label{eq:scalar1}
\end{align}
the diffeomorphism and the scalar constraints, respectively.

However, we will consider a redefinition of the shift and lapse functions. This redefinition makes the constraint algebra a true Lie algebra (see \cite{us} for details) and allows to complete the Dirac quantization. It is given by,
\begin{equation}\label{eq:class-NNx}
\tilde N^x:= N^x -2 N\frac{K_\varphi\sqrt{E^x}}{\left(E^x\right)'},\quad \tilde N :=- \frac{1}{E^\varphi}\left[N \frac{E^\varphi}{\left(E^x\right)'}\right]'.
\end{equation}
such that the total Hamiltonian is
\begin{equation}\label{eq:total-ham1}
H_T=\int dx (\tilde N\tilde H+\tilde N^xH_x),
\end{equation}
with
 \begin{equation}\label{eq:H_new-den}
\tilde H(\tilde N) :=\frac{1}{G}\int dx \tilde N
 \sqrt{E^x}E^\varphi\bigg[ K_\varphi^2-\frac{[(E^x)']^2}{4
 	(E^\varphi)^2}+\left(1-\frac{2 G M}{\sqrt{E^x}}\right)\bigg].
 \end{equation}
 Here, $M$ is the ADM mass, which, in absence of matter, completely determines solutions to the equations of motion and the constraints. Each constraint eliminates one phase space variable (per spacetime point). But one still needs to specify, if one wishes a fully gauge fixed theory, the radial coordinate and the spatial slicing (or equivalently lapse and shift functions). Nevertheless, different choices yield solutions that are diffeomorphically equivalent, and therefore their physical content is the same. 

When we set to define effective metrics, we will restrict ourselves to the set of solutions for which $\tilde N^x = 0$ and $\tilde N = 0$. This amounts to stationary slicings. In this situation, one can easily solve the theory and express the basic phase space variables in terms of two functional parameters $g(x),h(x)$ and the ADM mass observable, 
\begin{align}\nonumber
E^x(x) &= g(x),\quad (E^\varphi(x))^2 = \frac{\left[g'(x)\right]^2/4}{1+h^2(x) -\frac{2 G M}{\sqrt{g(x)}}}\\
K_x(x)&=\frac{\left[h'(x)\right]/2}{\sqrt{1+h(x) -\frac{2 G M}{\sqrt{g(x)}}}}, \quad K_{\varphi}(x) = h(x),
\end{align}
where $h(x)$ and $g(x)$ (such that $g(x)>0$ and $g'(x)\neq 0$) are arbitrary functions that represent the choice of coordinates for stationary spacetimes. Notice that it is common to make mass dependent coordinate changes (e.g. the ``tortoise'' coordinate) and in that case $g,h$ become functions of the ADM mass as well. Moreover, we require that the resulting spacetimes are asymptotically flat. This restricts $g(x)=x^2+{\cal O}(x^{-1})$ and $h(x)={\cal O}(x^{-1})$ in the limit $x\to\infty$. These conditions allow us to determine
\begin{equation}
N^2 = 1+h^2(x) -\frac{2 G M}{\sqrt{g(x)}},\quad N^x = 2\frac{h(x)\sqrt{g(x)}}{g'(x)}\sqrt{1+h^2(x) -\frac{2 G M}{\sqrt{g(x)}}},
\end{equation}
up to an irrelevant constant of integration for the lapse $N$ that is fixed by the condition $N(x)=1+{\cal O}(x^{-1})$ in the limit $x\to\infty$.

One can check that this identification of phase space variables 
is equivalent to imposing the gauge fixing conditions $\Phi_1=E^x(x)-g(x)$ and $\Phi_2=K_{\varphi}(x) - h(x)$. In many situations of interest, the gauge functions might depend on the canonical variables and $M$ as well, as it is the case, for instance, of the well-known Eddington--Finkelstein coordinates, discussed explicitly in Appendix \ref{app:class-EF}.

\section{Quantum theory: kinematics and improved dynamics}\label{sec:kin-impd}

Following \cite{us}, the basic mathematical building blocks of our quantum theory are $1$-dimensional oriented graphs, each containing a collection of consecutive edges $e_j$, each one ending in a vertex $v_j$.\footnote{In the standard literature \cite{us} one adopts the convention that the edge $e_j$ emerges from $v_j$. We find more convenient to adopt the opposite convention since the boundary conditions (due to the fact that we work with finite graphs) simplify considerably in the most quantum regions where the spin network ends using this convention.} The algebra of holonomies along the edges of these graphs (and the intertwiners on their vertices) provides the natural arena to construct the gravitational sector of the kinematical Hilbert space ${\cal H}^{\rm grav}_{\rm kin}$ of the theory, characterized by a basis of states $|\vec{k},\vec{\mu}\rangle$. Here, $k_j\in \mathbb{Z}$ and $\mu_j\in \mathbb{R}$ are valences of edges $e_j$ and vertices $v_j$, respectively. On this basis, kinematical operators corresponding to triads (and their spatial derivatives) are
\begin{equation}\label{eq:Edef}
  {\hat{E}^x(x) } |\vec{k},\vec{\nu}\rangle
  = \ell_{\rm Pl}^2 k_{j(x)} |\vec{k},\vec{\nu}\rangle,\quad   [\hat{E}^x(x)]' |\vec{k},\vec{\nu}\rangle
  = \ell_{\rm Pl}^2 (k_{j(x)+1}-k_{j(x)}) |\vec{k},\vec{\nu}\rangle,
\end{equation}
(notice that the derivative is with respect to a dimensionless coordinate)
where $j(x)$ is understood as the index corresponding to the edge $e_j$ going towards the vertex $v_j$ located at $x_j$, and 
\begin{equation}\label{eq:Ef-dist}
  \hat{E}^\varphi(x) |\vec{k},\vec{\mu}\rangle
  =\ell_{\rm Pl}^2 \sum_{\mu_j\in g} \delta(x-x_j)\mu_{j(x)} 
  |\vec{k},\vec{\mu}\rangle .
\end{equation}
Point holonomies \mbox{$\hat{{\cal N}}_{\rho_j}:=\widehat{\exp}(i\rho_j K_\varphi(x_j))$} of the connection $K_\varphi$ defined on a vertex $v_j$ have a well defined and simple  action on this single-vertex, state basis of ${\cal H}^{\rm grav}_{\rm kin}$. Concretely,
\begin{equation}\label{eq:Nmu-def}
  \hat{{\cal N}}_{\rho_j}|\mu_j\rangle = |\mu_j+\rho_j\rangle.
\end{equation}
There are also well-defined operators corresponding to holonomies of the connection component $K_x$. However, in the Abelian Hamiltonian constraint there are no components of the curvature proportional to $K_x$. Thus, it is not necessary to construct explicitly the operator corresponding to the holonomy along edges $e_j$.

We will now consider the {\it improved dynamics } scheme introduced by Chiou {\em et al.} \cite{chiou}. For spherically symmetric spacetimes on homogeneous slicings there are several schemes proposed in the literature \cite{ab,lm,ck,bv,dc,cgp,cs,oss,cctr,yks,js,chiouBH,bkd,djs,bmm,aos}. For inhomogeneous Gowdy models with local rotational symmetry these ideas have also been studied in detail in Ref. \cite{gowdy-lrs}.
We will adopt the main aspects of this quantization and adapt it to black hole spacetimes.  

The technical implementation of this scheme starts with the components of the classical curvature (of the real connection) approximated by holonomies of finite closed loops along suitable edges generated by the Killing vectors, such that the physical area enclosed by these plaquettes equals the $1$st nonzero eigenvalue of the full LQG area operator (known as the \emph{area gap}), and denoted by $\Delta$.

In the Abelian Hamiltonian constraint there remains only one component of the curvature, there is no dependence on $K_x$. The closed holonomy that will approximate the remaining component of the curvature in the quantum theory on each vertex $v_j$ is obtained by considering a plaquette adapted to a 2-sphere, enclosing a physical area
\begin{equation}\label{eq:area-cond}
4\pi \ell^2_{\rm Pl}k_j \bar\rho^2_j = \Delta,
\end{equation}
where $\ell^2_{\rm Pl}k_j$ is the eigenvalue of the kinematical operator $\hat E^x(x_j)$, defined in Eq. \eqref{eq:Edef}. Now, point holonomies \eqref{eq:Nmu-def} of fractional length $\bar\rho_j$ will produce a shift in a state $|\mu_j\rangle$ which depends on the spectrum of some kinematical operators. Concretely, $|\mu_j\rangle\to|\mu_j+\bar\rho_j\rangle $, and given the above relation,
\begin{equation}
\bar\rho_j = \sqrt{\frac{\Delta}{4\pi \ell^2_{\rm Pl}k_j}}.
\end{equation}
Therefore, it will be convenient to adopt a more appropriate state labeling $|\nu_{j}\rangle$ with $\nu_{j}=\sqrt{k_j}\mu_{j}/\lambda$, and $\lambda^2=\Delta/4\pi \ell_{\rm Pl}^2$. 
Point holonomies of the form \mbox{$\hat{{\cal N}}_{\bar\rho_j}:=\widehat{\exp}(i\bar\rho_j K_\varphi(x_j))$} again have a well-defined and simple  action on this new (single-vertex) state basis of ${\cal H}^{\rm grav}_{\rm kin}$
\begin{equation}\label{eq:Ndef}
  \hat{{\cal N}}_{\bar\rho_j}|\nu_j\rangle = |\nu_j+1\rangle.
\end{equation}

For further details, see Ref. \cite{gowdy-lrs}. The basis of states is now denoted by $|\vec{k},\vec{\nu}\rangle$, and its elements are normalized to $\langle\vec{k},\vec{\nu}|\vec{k}',\vec{\nu}'\rangle=\delta_{\vec{k}\vec{k}'}\delta_{\vec{\nu}\vec{\nu}'}$ in ${\cal H}^{\rm grav}_{\rm kin}$. 

On this basis, the set of basic kinematical operators defined above in Eq. \eqref{eq:Edef} remains the same. However, it is more convenient to replace the kinematical operator $ \hat{E}^\varphi(x)$, defined in Eq. \eqref{eq:Ef-dist}, by the volume operator density
\begin{equation}\label{eq:V-dist}
  \hat{V}(x) |\vec{k},\vec{\nu}\rangle
  = \lambda\ell_{\rm Pl}^3 \sum_{\nu_j\in g} \delta(x-x_j)\nu_{j(x)} 
  |\vec{k},\vec{\nu}\rangle .
\end{equation}

In total, a basis in ${\cal H}_{\rm kin}$ is given by $|\vec{k},\vec{\nu},M\rangle$ with norm
\begin{equation}
\langle\vec{k},\vec{\nu},M|\vec{k}',\vec{\nu}',M'\rangle=\delta_{\vec{k}\vec{k}'}\delta_{\vec{\nu}\vec{\nu}'}\delta(M-M'),
\end{equation} 
where we adopt a standard representation for $\hat M$ (and its conjugate variable $\hat P$).\footnote{Note that the Abelian Hamiltonian constraint is independent of $P$, it only depends on $M$.}

The quantum operator corresponding to the scalar constraint is
 \begin{equation}\label{eq:H_new-den1}
\hat H(\tilde N) :=\sum_j \tilde N(x_j)
 \sqrt{\hat E^x(x_j)}\hat E^\varphi(x_j)\bigg[\hat\Theta(x_j)-\frac{[(\hat E^x(x_j))']^2}{4
 	[\hat E^\varphi(x_j)]^2}+\left(1-\frac{2 G \hat M}{\sqrt{\hat E^x(x_j)}}\right)\bigg],
 \end{equation}
where we define
\begin{equation}
\hat\Theta(x_j)= \frac{1}{[\hat E^\varphi(x_j)]}\widehat{\frac{\sin\left(\bar\rho_jK_\varphi(x_j)\right)}{\bar\rho_j}}\hat E^\varphi(x_j)  \widehat{\frac{\sin\left(\bar\rho_jK_\varphi(x_j)\right)}{\bar\rho_j}}
\end{equation}
and
\begin{equation}
\frac{1}{[\hat E^\varphi(x_j)]}:=\frac{\sqrt{k_j}}{\lambda\ell_{\rm Pl}^2 \nu_{j}},
\end{equation}
i.e., a scalar version of the inverse of the triad density of Eq. \eqref{eq:Ef-dist} in the $\nu$-representation. The operator $\hat H(\tilde N)$ has a well-defined action on ${\cal H}_{\rm kin}$.\footnote{Subtleties could arise when $\nu_{j(x)}=0$ in some vertex. Notice that at the kinematical level this problem may be avoided by choosing superselection sectors for $\mu_j$ that do not include $0$. See for instance Refs. \cite{us,gowdy-lrs,qsd,abl} for alternative treatments of this issue. On the physical Hilbert space this issue will not affect our discussion  since the physical operator representing $[\hat E^\varphi(x_j)]^2$ restricted to this physical sector has a spectrum with no vanishing eigenvalues --see Eq \eqref{eq:hephi} below.} It acts on each vertex of the kinematical states $|\vec k,\vec \nu,M\rangle$ as a difference operator with support in lattices ${\cal L}_{\epsilon_j}$ of the label $\nu_j$ such that $\nu_j=\epsilon_j+2m_j$ with $\epsilon_j\in[0,2]$ a collection of continuous parameters that label each $\nu$-lattice and $m_j\in \mathbb{Z}$ a collection of integers, one for each vertex $v_j$. Besides, the Hamiltonian constraint operator is diagonal with respect to the labels $k_j$ and $M$.
 
In order to simplify the study of the solutions to this Hamiltonian constraint, we will follow the ideas of \cite{slqc}. Namely, on each vertex $v_j$, we will restrict the analysis to the lattice ${\cal L}_{\epsilon_j=0}$ (such that $\nu_j=2m_j$ and $m_j\neq 0$ for all $j$),\footnote{We expect that, as in LQC, the restriction to other lattices will provide qualitatively similar results, especially for the semiclassical states we are interested in here.} and change the representation to the one in which the trigonometric functions of $K_\varphi(x_j)$ act by multiplication, by means of a suitable Fourier transform (see Ref. \cite{slqc} for details). This is the representation originally adopted in Ref. \cite{us}. Then, one can easily identify physical states (out of an exact integration of the solutions to the Hamiltonian constraint) and observables, and then construct the physical Hilbert space of the theory. We will provide further details in the next section.

\section{Physical Hilbert space and observables}\label{sec:phys}

It is then not difficult to follow the quantization procedure of \cite{us} and identify suitable physical states and parametrized observables. More concretely, a basis of physical states is provided by $|M,\vec k\rangle$, normalized to $\langle M,\vec k|M',\vec k'\rangle=\delta(M-M')\delta_{\vec k,\vec k'}$.

The basic relevant observables are the mass $\hat M$, which has a well-defined action on physical states
\begin{equation}
\hat M |M,\vec k\rangle = M |M,\vec k\rangle,
\end{equation}
and the collection of observables associated with $\vec k$, which can be defined as a parametrized observable (following similar arguments as those of Ref. \cite{us}), namely,
\begin{equation}\label{eq:ObsO}
\hat O(z)|M,\vec k\rangle=\ell_{\rm Pl}^2 k_{{\rm Int}(Sz)}|M,\vec k\rangle,
\end{equation}
with $S$ the total number of vertices, $z\in [0, 1]$ is a continuous parameter that allows us to label the action of observables on states with any (finite) number of vertices and ${\rm Int}(Sz)$ means the integer part of $S z$. The physical meaning of this observable is simple: it just codifies the (quantized) areas of the spheres of symmetry.

In what follows, we will work with spin networks with a finite but large number of vertices $S$. For simplicity, we restrict the study to spin networks whose values of $k_j$  are associated with a lattice with equidistant spacing such that, 
\begin{equation}
x_j = \delta x(j+j_0),
\end{equation} 
where $j_0\geq 1$ is an integer that will be specified below and $\delta x$ is the step of the lattice of the coordinate $x$ that we choose to be $\delta x=\ell_{\rm Pl}$ (other choices of $\delta x$ are allowed provided that $\frac{\ell_{\rm Pl}}{2\sqrt{k_0}}<\delta x < \ell_{\rm Pl}\sqrt{k_0}$). This choice amounts to choose the function $z(x)$ as $z(x)=x/(S\delta x)$, such that $z(x_j)=(j+j_0)/S$.

For instance, in this family of states, the triad $E^x$ and its spatial derivative can be easily represented as physical parametrized observables as 
\begin{align}\label{eq:hex}
  &\hat E^x(x_j)|M,\vec k\rangle=\hat O(z(x_j))|M,\vec k\rangle=\ell_{\rm Pl}^2 k_{j}|M,\vec k\rangle=x^2_{j}|M,\vec k\rangle.\\\label{eq:hdex}
  & [\hat E^x(x_j)]'|M,\vec k\rangle=\frac{(x_j+\delta x)^2-x_j^2}{\delta x^2}|M,\vec k\rangle=(2x_j+\delta x)|M,\vec k\rangle.
\end{align} 

Using the Hamiltonian constraint (\ref{eq:H_new-den}), the (square of the) triad $E^\varphi$ is represented by
\begin{equation}
\label{eq:hephi}
(\hat E^\varphi(x_j))^2 = \frac{\left[(\hat E^x(x_j))'\right]^2/4}{1+
\widehat{\frac{{\sin^2\left(\bar\rho_j K_\varphi(x_j)\right)}}{{\bar\rho}_j^2}} -\frac{2 G \hat M}{\sqrt{|\hat E^x(x_j)|}}},
\end{equation}
where $K_\varphi(x_j)$ can depend on $\hat M$ or $\hat O(z)$. For instance, for Eddington--Finkelstein coordinates, consider (\ref{eq:hex}) in conjunction with (A1) in the appendix. 
This yields, 
\begin{equation}
\widehat{\frac{{\sin^2\left({\bar\rho}_j K_\varphi(x_j)\right)}}{{\bar\rho}_j^2}}=\frac{(2G\hat M)^2}{{\hat O(z(x_j))}}\frac{1}{{1+\frac{2G\hat M}{\sqrt{\hat O(z(x_j))}}}}.
\end{equation}

Let us notice that a minimum requirement for $\hat{E}^\varphi$ to be a well defined self-adjoint operator is that, in terms of eigenvalues, we have,
\begin{equation}\label{eq:imp}
  1+\frac{\sin^2\left(\bar\rho_j K_\varphi(x_j)\right)}{\bar\rho^2_j} -\frac{2 G M}{\sqrt{E^x(x_j)}}>0, \quad \forall x_j,M. 
\end{equation}

This condition leads to a minimum eigenvalue of $E^x(x)$, $\ell_{\rm Pl}^2 k_0$, and at this point the curvature is maximum.
Let us study in detail this situation. This implies both $\sin\left(\bar\rho_j K_\varphi(x_j)\right)=1$, and $\bar\rho_j$ given by  \eqref{eq:area-cond}.  For a given mass $M$, the smallest area of the 2-spheres must be such that
\begin{equation}\label{eq:k0-cond}
  \left(1+\frac{4\pi \ell_{\rm Pl}^2 k_0}{\Delta}\right) -\frac{2 G M}{\sqrt{\ell_{\rm Pl}^2 k_0}}>0.
\end{equation}
Assuming that $k_0\gg 1$, we get
\begin{equation}
  k_0 > \left(\frac{2 G M \Delta}{4\pi \ell_{\rm Pl}^3}\right)^{2/3} = \tilde k_0.
\end{equation}
Note that, since $\Delta \simeq \ell_{\rm Pl}^2$, the limit $k_0\gg 1$ implies $M\gg m_{\rm Pl}$ (this corresponds to large black holes compared to the Planck mass). Let us consider the first integer $k_0$ that is larger than $\tilde k_0$. For states with $\tilde k_0\gg 1$, the minimum value of the smallest 2-sphere is\footnote{This scaling with the mass is in agreement with the prescription proposed by Ashtekar, Olmedo and Singh \cite{aos}.}
\begin{equation}
  k_0 \simeq\tilde k_0 \propto M^{2/3}.
\end{equation}

Therefore, we restrict the domain of definition of the Hamiltonian constraint operator to states satisfying the conditions above, namely such that $\hat E^x$ has a minimum eigenvalue $\ell_{\rm Pl}^2 k_0$ determined by the mass $M$ through $\tilde k_0$. For states that involve superpositions of the mass, this last condition is highly nontrivial since one would need to superpose several spin networks. But, as we will see below, we can restrict the study to a family of physical semiclassical states (actually it can be easily generalized to other families of semiclassical states) with support in a single spin network. 

In general one cannot restrict the domain of operators as we are doing. This is possible here because we are in the physical space of states and there is no physical observable that connects different spin networks. Notice that the restriction of the domain eliminates the singularity. This makes possible to continue the manifold beyond where the classical singularity used to be. We have discussed this in the context of the previous quantization \cite{us}, but we will not concentrate on this point in this paper.

 In general, once a choice for $K_\varphi$ is made, in the classical theory it amounts to a particular space-time slicing (we restrict ourselves to stationary, asymptotically spatially flat ones). In turn, this specifies the lapse and shift functions that appear in Eq. \eqref{eq:class-NNx} ---we must recall that for the type of slicings under consideration $\tilde N^x=0=\tilde N$. However, the classical expressions must be promoted to quantum operators. Specifically, we choose
\begin{equation}\label{eq:q-lapshi}
  \hat N^2(x_j) :=\frac{1}{4}\frac{([\hat E^x(x_j)]')^2}{(\hat E^\varphi(x_j))^2},\quad
  \hat N^x(x_j) =\sqrt{\frac{\hat E^x(x_j)}{(\hat E^\varphi(x_j))^2}}\widehat{\frac{\sin\left(2{\bar\rho}_j K_\varphi(x_j)\right)}{2{\bar\rho}_j}}.
\end{equation}
Let us briefly comment that the classical expression for the shift involves components of the curvature $K_\varphi$ that must be polymerized. The choice we make is motivated as follows. In the expression for the shift, components of the curvature do not appear all squared, as in the Hamiltonian constraint. Then, if one adopts the same polymerization, namely $K_\varphi\to\sin(\rho K_\varphi)/\rho$ for it, the two kinematical operators $\widehat{\sin(\rho K_\varphi)/\rho}$ and $\widehat{\sin^2(\rho K_\varphi)/\rho^2}$ will not share the same lattices in $\nu_j$ (they are defined on lattices of step one and two, respectively).\footnote{One could actually consider the restriction of the operator $\widehat{\sin(\rho K_\varphi)/\rho}$ to the domain (lattices) of its square, but this requires extra consideration. It is for this reasson that we adopt a simpler option.} However, if the polymerization of the scalar constraint is given by  $\widehat{\sin^2(\rho K_\varphi)/\rho^2}$, the simplest polymerization for the shift that is compatible with its $\nu$-lattices is given by  $\widehat{\sin(2\rho K_\varphi)/2\rho}$. Hence, this operator will be well-defined on physical states provided by the scalar constraint (see Ref. \cite{us} for further details). \footnote{There is a parallelism in homogeneous LQC when one constructs a physical operator related to the Hubble rate. In Ref. \cite{presc-lqc} it was chosen to leave invariant the superselection sectors related to the Hamiltonian constraint.}

Physical states, in general, are superpositions of mass and spin networks (1D lattices). However, we will restrict the study to a family of physical states that are sharply peaked in the mass and a concrete spin network for simplicity. This restricts the expectation values and dispersions of the mass, since they must be compatible with condition \eqref{eq:k0-cond} and the restriction to a single spin network. This implies that the state in the mass must have support in an interval $M\in[M_0-\delta M_0,M_0+\delta M_0]$, for some $M_0$, such that $k_0(M_0+\delta M_0)=k_0(M_0-\delta M_0)=k_0(M_0)$ with 
\begin{equation}
k_0(M_0)={\rm Int}\left[\left(\frac{2 G M_0 \Delta}{4\pi \ell_{\rm Pl}^3}\right)^{2/3}\right].
\end{equation}
One can easily see that in the limit $M_0\gg m_{\rm Pl}$ this implies 
\begin{equation}
\delta M_0 \leq \frac{3}{2}\left(\frac{4\pi \ell_{\rm Pl}^3}{2G\Delta}\right)^{2/3}M_0^{1/3}=\widetilde{\Delta M_0}.
\end{equation}

These nontrivial conditions are met, for instance, by the states 
\begin{equation}\label{eq:psi}
|\psi\rangle=\frac{1}{\Delta M_0}\int dM e^{iM P_0/\hbar}\cos\left[\frac{\pi(M-M_0)}{2\Delta M_0}\right]\Theta(M-M_0+\Delta M_0)\Theta(M_0+\Delta M_0-M)|M,k_S,\ldots,k_0\rangle
\end{equation}
with $M_0\gg m_{\rm Pl}$ and $\Delta M_0 \leq \widetilde{\Delta M_0}$.
\footnote{If $\Delta M_0$ is not sufficiently small, one should consider suitable superpositions on different spin networks, each one respecting condition \eqref{eq:k0-cond}.} On these states, the observable $\hat M$ and its conjugate variable $\hat P$ (such that $[\hat M,\hat P]=i\hbar$) satisfy
\begin{align}\nonumber
  &\langle \hat M \rangle = M_0,\quad \Delta M= \Delta M_0\sqrt{ \frac{1}{3} - \frac{2 }{\pi^{2}} },\\
  &\langle \hat P \rangle = P_0,\quad \Delta P= \frac{\hbar\pi}{2\Delta M_0}.
\end{align}
These states fulfill $\Delta M \Delta P=\pi\hbar/2$. Relative dispersions will be small if $M_0\gg \Delta M_0$ and $P_0\gg \hbar/(2\Delta M_0)$ since
\begin{equation}\label{eq:deltam}
\Delta M_0 \leq \frac{3}{2}\left(\frac{4\pi \ell_{\rm Pl}^3}{2G\Delta}\right)^{2/3}M_0^{1/3}.
\end{equation}
In summary, this choice guaranties that the support of the state (i.e. $M\in[M_0-\Delta M_0,M_0+\Delta M_0])$ requires only a single spin network compatible with \eqref{eq:k0-cond} since $k_0(M_0+\Delta M_0)=k_0(M_0-\Delta M_0)=k_0(M_0)$. 

\section{Effective metric}\label{sec:efft}

In order to further explore the physical consequences of this prescription, let us construct the line element of the spacetime as follows. We adopt an Eddington--Finkelstein, horizon penetrating slicing determined by the condition we considered before,
\begin{equation}\label{eq:gauge}
\frac{\widehat{\sin^2\left({\bar\rho}_j K_\varphi(x_j)\right)}}{{\bar\rho}_j^2}=\frac{4G^2\hat M^2}{{\hat E^x(x_j)}}\frac{1}{{1+\frac{2G\hat M}{\sqrt{\hat E^x(x_j)}}}}.
\end{equation}

Then, one can easily construct the operators corresponding to the components of the spacetime metric out of Eqs. \eqref{eq:hex}, \eqref{eq:hdex}, \eqref{eq:hephi} and \eqref{eq:q-lapshi}. They  are given by
\begin{eqnarray}\nonumber
\hat g_{tt} &=& -\left(1-\frac{\hat r_S}{\sqrt{\hat E^x}}+\frac{\Delta}{4\pi}\frac{\hat r_S^4 }{(\hat E^x)^3\left(1+\frac{\hat r_S}{\sqrt{\hat E^x}}\right)^2}\right),\quad  \hat g_{xx} = \frac{\left\{\widehat{\left[E^x\right]'}\right\}^2}{4 \hat E^x}\left(1+\frac{\hat r_S}{\sqrt{\hat E^x}}\right),\\
\hat g_{tx} &=& - \frac{\hat r_S \widehat{\left[E^x\right]'}}{2 \hat E^x}\sqrt{1-\frac{\Delta}{4\pi}\frac{\hat r_S^2 }{(\hat E^x)^2\left(1+\frac{\hat r_S}{\sqrt{\hat E^x}}\right)}}, \quad g_{\theta\theta}=\hat E^x,\quad g_{\phi\phi}=\hat E^x\sin^2\theta,\label{eq:hatgmunu}
\end{eqnarray}
after replacing condition \eqref{eq:gauge}. Here, $\hat r_S = 2G\hat M$. The effective metric is defined as $g_{\mu\nu}=\langle \hat g_{\mu\nu}\rangle$, where the expectation value is computed on the state \eqref{eq:psi}. \footnote{The effective metric can be obtained by other means. For instance, using quantum field theories on these quantum spacetimes and deriving the corresponding dressed geometry or defining an appropriate Riemann curvature quantum operator, computing the expectation values of its components and finally reading the resulting effective metric. In these cases, one should expect qualitatively similar but quantitatively different results. Moreover, further corrections will appear in those situations where superpositions of spin networks become relevant.} 

Quantum effects on this effective geometries are present in i) the cutoff in the expectation value $\hat E^x(x)$, ii) the discreteness inherited  by $\widehat{\left[E^x(x)\right]'}$, iii) polymer corrections due to the representation of curvature components, and iv) superpositions in the  mass $\hat M$. However, we will focus here in the effects due to i) - iii), since they are the most prominent ones. The derivation of this effective metric and a discussion about the subleading effects due to iv) can be found in Appendix \ref{app2}. Hence, in the limit in which $\Delta M_0$ is small, we can write the effective metric as $g_{\mu\nu}={}^{(0)}g_{\mu\nu}+{}^{(2)}g_{\mu\nu}\Delta M_0^2+\ldots$ We will focus our analysis on  ${}^{(0)}g_{\mu\nu}$ in the following. For details about the subleading contributions ${}^{(2)}g_{\mu\nu}$, see Appendix \ref{app2}. The lowest order is explicitly given by
\begin{align}\nonumber
&{}^{(0)}ds^2:={}^{(0)}g_{\mu\nu}dx^\mu dx^\nu=-\left(1-\frac{r_S}{x+x_0}+\frac{\Delta}{4\pi}\frac{r_S^4 }{(x+x_0)^6\left(1+\frac{r_S}{x+x_0}\right)^2}\right)dt^2\\ \nonumber
&-\frac{r_S}{(x+x_0)}\left(1+\frac{\delta x}{2(x+x_0)}\right)\left(\sqrt{1-\frac{\Delta}{4\pi}\frac{r_S^2 }{(x+x_0)^4\left(1+\frac{r_S}{x+x_0}\right)}}\right)dtdx\\
&+\left(1+\frac{r_S}{x+x_0}\right)\left(1+\frac{\delta x}{2(x+x_0)}\right)^2dx^2+(x+x_0)^2d\omega^2. \label{eq:g0}
\end{align}
Here $d\omega^2$ is again the line element of the unit 2-sphere. The effective metric depends on two parameters, $M_0$ and $\delta x$ (since both $r_S=2GM_0$ and $x_0=\left(\frac{2 G M_0 \Delta}{4\pi}\right)^{1/3}$ are determined by $M_0$). $\Delta$ is the area gap parameter that is determined by the full theory. Here, for simplicity, we take the continuum limit $j\to x/\delta x \in \mathbb{R}^+$ (see Appendix \ref{app:cont} for details) keeping $\delta x = \ell_{\rm Pl} $ nonvanishing.\footnote{Other choices with $\delta x \in[ \ell_{\rm Pl}^2/(2x_0) ,x_0]$ are available. Note however that $\delta x<\ell^2_{\rm Pl}/(2x_0)$ is forbidden, otherwise the area gap of the 2-spheres determined by the spectrum of $\hat E^x$ would be smaller than $\ell^2_{\rm Pl}$. Besides, $\delta x>x_0$ is not allowed since this would involve $j_0<1$.}

The main properties of these effective geometries, which we will discuss in detail soon, can be summarized in:  
\begin{itemize}
\item The effective spacetime metric has a Killing vector field $X^a$ that is time-like (space-like) outside (inside) the horizon. One can easily see that the latter is located at ${}^{(0)}g_{00}(x_H)=0$.
\item ${}^{(0)}g_{\mu\nu}$ agrees very well with the classical metric until $x\simeq x_0$. A description by means of a continuum effective metric is valid until $x\simeq \delta x$.
\item Curvature scalars (Kretschmann, Ricci scalar and Ricci contracted with itself) are bounded above, and in the limit $M_0\gg m_{\rm Pl}$ the upper bounds are independent of $M_0$ and $\delta x$. 
\item Quantum corrections can be characterized as an effective stress-energy tensor (via the nonvanishing Einstein tensor), characterized by an effective energy density, radial and tangential pressures. It violates the null energy condition.  
\item For macroscopic black holes the Komar and Misner--Sharpe masses agree at spatial infinity with the ADM mass. But in the most quantum region they differ from $M_0$. 
\end{itemize}

Therefore, these effective metrics solve the classical singularity at high curvatures, and agree very well with classical GR in the low curvature regime. This description is valid for all black holes, including those with masses as small as $M_0\simeq 10^3$. \footnote{For smaller masses, one should solve exactly Eq. \eqref{eq:k0-cond}. Besides, effects due to the discreteness of quantum geometry and fluctuations of both the mass and the quantum geometry are expected to be important and must be analyzed all together.}

Interestingly, the classical limit corresponds to $\hbar \to 0$, which amounts to $\Delta \to 0$ (and therefore $x_0\to 0$) and $\delta x\to 0$ (if in addition one considers corrections from fluctuations in the mass this classical limit implies also $\Delta M_0\to 0$). The resulting metric has a vanishing Ricci tensor. Thus, by Birkhoff's theorem, the effective metric in the exterior region is locally diffeomorphic to the Schwarzschild metric. The extension of this metric to the interior finds the usual future singularity in the strong curvature region.

\subsection{Curvature of the effective spacetime}

\begin{figure}[h]
{\centering     
  \includegraphics[width = 0.49\textwidth]{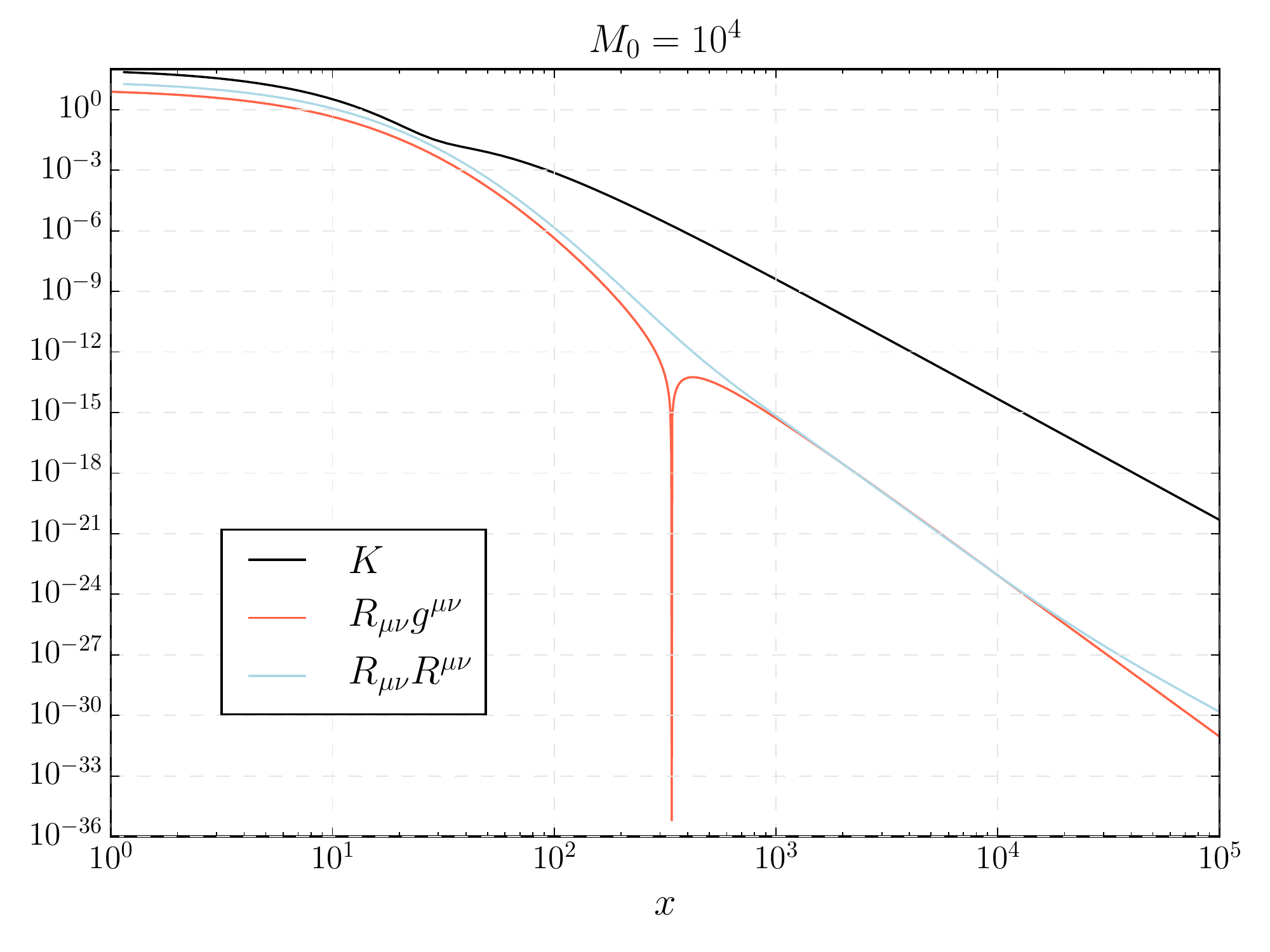}
\includegraphics[width = 0.49\textwidth]{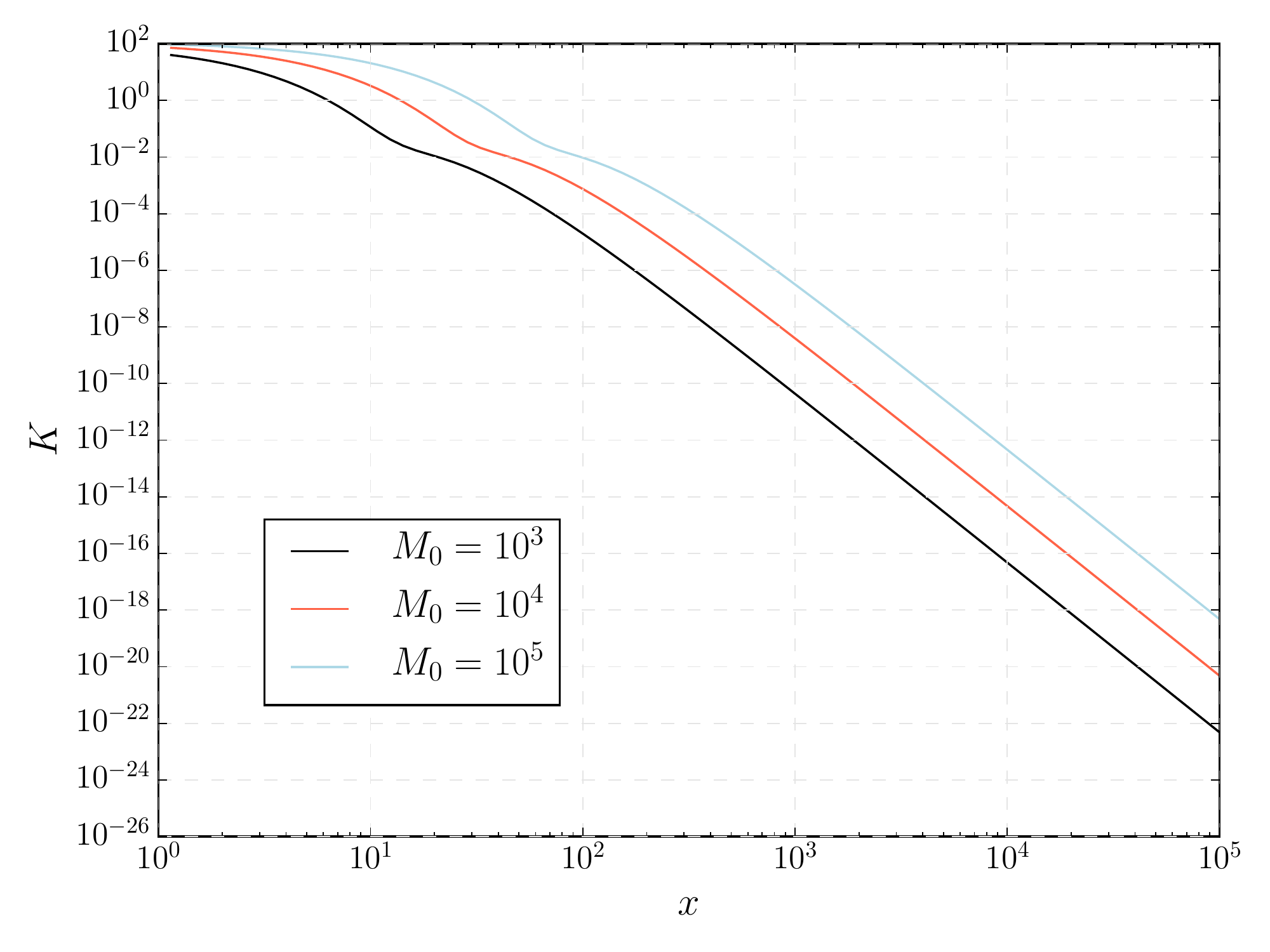}
}
\caption{Left Panel: Curvature invariants for a choice of the mass parameter corresponding to $M_0=10^4$ and $\delta x=1$. Right Panel: Kretschmann scalar $K$ for different values of the mass $M_0$ and $\delta x=1$.}
\label{fig:k}
\end{figure}

\begin{figure}[h]
{\centering     
  \includegraphics[width = 0.49\textwidth]{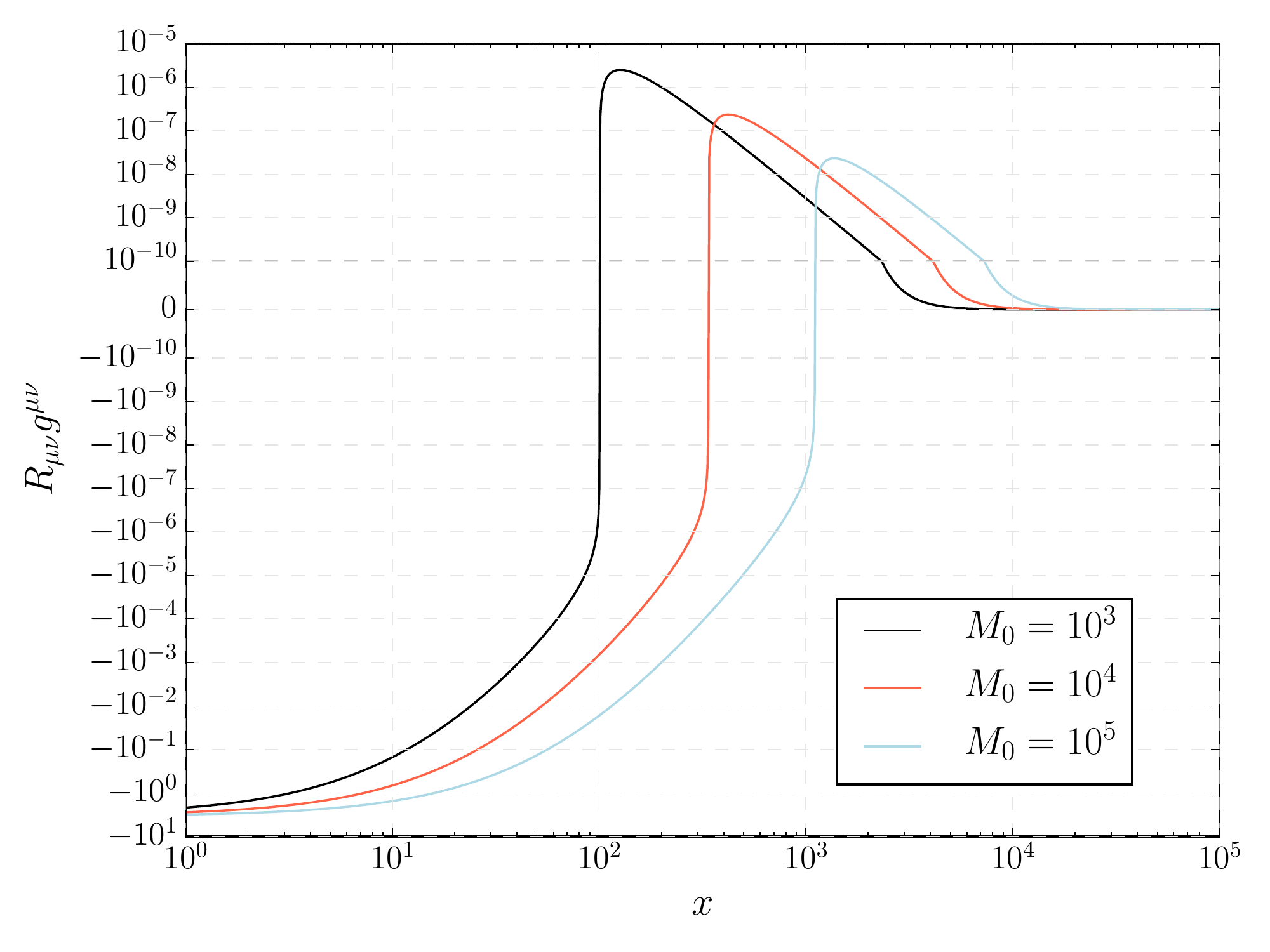}
  \includegraphics[width = 0.49\textwidth]{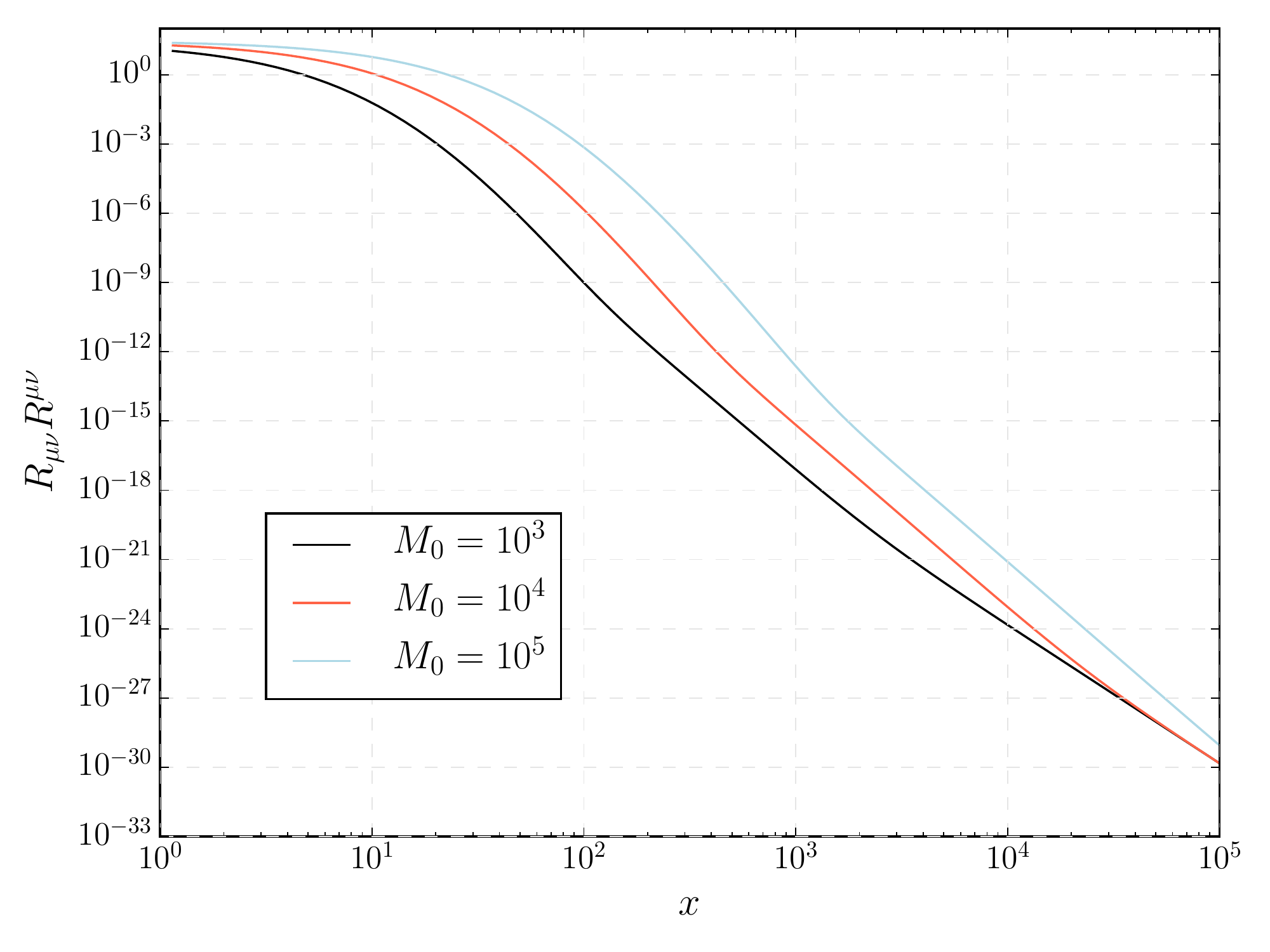}  
}
\caption{Ricci scalar $R_{\mu\nu}g^{\mu\nu}$ (left panel) and Ricci tensor squared $R_{\mu\nu}R^{\mu\nu}$ (right panel) for different values of the mass $M_0$ and $\delta x=1$. }
\label{fig:r2}
\end{figure}

Let us now analyze several properties of the curvature of the effective metric ${}^{(0)}g_{\mu\nu}$. We consider the Kretschmann scalar $K=R_{\mu\nu\rho\sigma}R^{\mu\nu\rho\sigma}$, the Ricci tensor squared $R_{\mu\nu}R^{\mu\nu}$ and the Ricci scalar $R_{\mu\nu}g^{\mu\nu}$ (Weyl scalar can be easily obtained out of these three scalars). We have found the following asymptotic expressions at spatial infinity
\begin{align}\nonumber
    K &= \frac{6(8 G^2 M_0^2 +4GM_0\delta x+\delta x^2)}{x^6}+{\cal O}(x^{-7}),\\ \label{eq: asymp-curv}
    R_{\mu\nu}g^{\mu\nu}&=\frac{3\delta x(2 G M_0 +\delta x)}{2x^4}+{\cal O}(x^{-10}),\quad R_{\mu\nu}R^{\mu\nu} = \frac{3\delta x^2}{2x^6}+{\cal O}(x^{-7}).
\end{align}
As we see, the main deviations from classical GR are dominated by the parameter $\delta x$. On the other hand, in the most quantum region ($x\ll x_0$) and in the limit $M_0\gg m_{\rm Pl}$ (and for those situations where $\delta x$ is independent of $M_0$), we obtain
\begin{align} \label{eq: q-curv}
    K &= \frac{5760\pi^2}{\Delta^2}+{\cal O}(M_0^{-1/3}), \quad
    R_{\mu\nu}g^{\mu\nu}=-\frac{24\pi}{\Delta}+{\cal O}(M_0^{-1/3}),\quad R_{\mu\nu}R^{\mu\nu} = \frac{1440\pi^2}{\Delta^2}+{\cal O}(M_0^{-1/3}).
\end{align}
As we see, in the most quantum region, macroscopic black holes show curvature invariants that reach upper bounds independent to $M_0$ and are fully determined by the area gap $\Delta$. This is in contrast to previous treatments where the upper bounds grew with $M_0$ and become trans-Planckian for large black holes. In Figs. \ref{fig:k} and \ref{fig:r2} we show these scalars including the most quantum region for different values of $M_0$. In the following, we adopt Planck units. Then, $\Delta=4\pi\sqrt{3}\gamma$ as it is usual in LQC (recall that here we adopt $\gamma=1$). Besides, we adopt here $\delta x=1$.
As we see, all curvature scalars reach a similar Planckian magnitude. The scalars constructed with the Ricci tensor decrease fast away from the quantum region, at least as fast as the Krestchmann scalar, in agreement with the analytical expressions deduced in Eq. \eqref{eq: asymp-curv}. Besides, we see that the Ricci scalar is negative and Planck order in the most quantum region. It decreases and switches sign as one moves towards low curvature regions. This is a consequence of the quantum corrections in these effective geometries and their balance in the different regions of the space-time. We will discuss them in more detail in the next section.

\subsection{Effective stress-energy tensor}

The novel properties of the effective quantum geometries ${}^{(0)}g_{\mu\nu}$ can be codified into an effective stress-energy tensor 
\begin{equation}
T_{\mu\nu}:=\frac{1}{8\pi G} G_{\mu\nu},
\end{equation}
where $G_{\mu\nu}$ is the Einstein tensor. In turn, $T_{\mu\nu}$ is characterized by an effective energy density $\rho$ and radial and tangential pressures densities, $p_x$ and $p_{||}$, respectively. 

In the exterior region, namely, the region where ${}^{(0)}g_{\mu\nu}$ has a Killing vector field $X^\mu$ that is time-like, the components of the stress-energy tensor  are defined by means of
\begin{equation}
\rho := T_{\mu\nu}\frac{X^\mu X^\nu}{(-X^\rho X_\rho)},
\end{equation}
\begin{equation}
p_x := T_{\mu\nu}\frac{r^\mu r^\nu}{r^\rho r_\rho},
\end{equation}
and 
\begin{equation}
p_{||} := T_{\mu\nu}\frac{\theta^\mu \theta^\nu}{\theta^\rho \theta_\rho},
\end{equation}
where, in addition, $r^\mu$ and $\theta^\mu$ are vector fields pointing in the radial and $\theta$-angular direction, respectively. In the interior region, on the other hand, $X^\mu$ becomes space-like while $r^\mu$ is now time like. Therefore, in this situation, the previous expressions are still valid but now $r^\mu$ plays the role of $X^\mu$ (and viceversa).

The asymptotic behavior  of these quantities at $x\to\infty$ can be easily derived
\begin{align}\nonumber
    \rho &=\frac{1}{8\pi G}\frac{\delta x(8 G M_0 +3\delta x)}{4x^4}+{\cal O}(x^{-5}),\quad p_x = -\frac{1}{8\pi G}\frac{\delta x}{x^3}+{\cal O}(x^{-4}) ,\quad p_{||} = \frac{1}{8\pi G}\frac{\delta x}{2x^3}+{\cal O}(x^{-4}).
\end{align}
This shows that the fall-off of the effective stress-energy tensor is sufficiently fast. Hence, the effective metric will approach at spatial infinity the Minkowski metric in the $(t,x,\theta,\phi)$ coordinates with the standard fall-off conditions \cite{geroch-cinn,ah,aa-ein}. In addition, the asymptotic behavior of the stress-energy tensor is dominated by the parameter $\delta x$ (corrections due to the area gap $\Delta$ of LQG are subdominant). 
On the other hand, in the most quantum region ($x\ll x_0$), and in the limit $M_0\gg m_{\rm Pl}$, we have
\begin{align}\nonumber
    \rho &=\frac{1}{8\pi G}\frac{12 \pi}{\Delta}+{\cal O}(M_0^{-1/3}),\quad p_x = -\frac{1}{8\pi G}\frac{12 \pi}{\Delta}+{\cal O}(M_0^{-1/3}) ,\quad p_{||} = \frac{1}{8\pi G}\frac{24 \pi}{\Delta}+{\cal O}(M_0^{-1/3}).
\end{align}
As we see, components of the stress-energy tensor reach upper bounds that, for macroscopic black holes, become universal and completely specified by the area gap $\Delta$ of LQG.

\begin{figure}[h]
{\centering     
  \includegraphics[width = 0.49\textwidth]{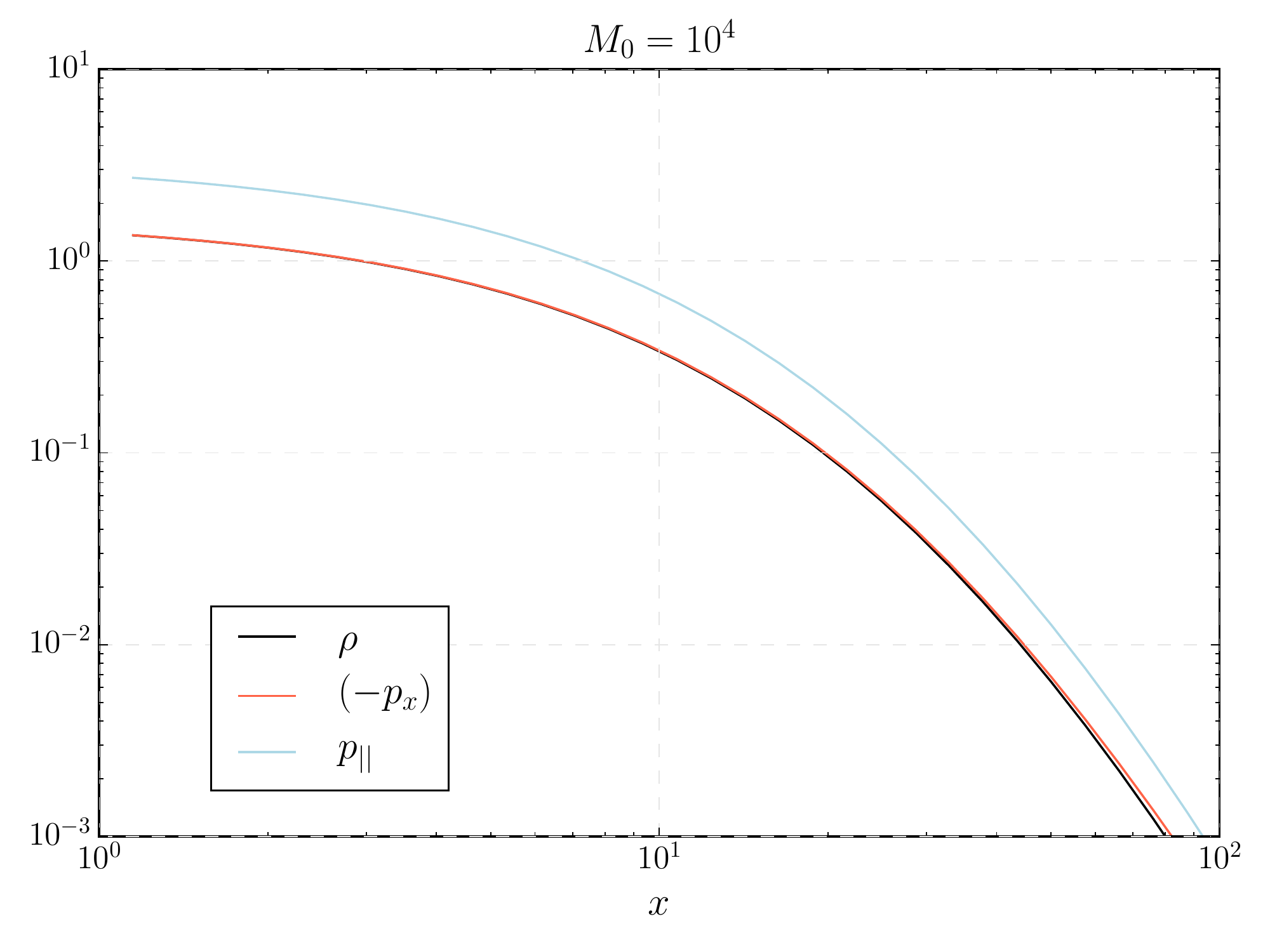}
  \includegraphics[width = 0.49\textwidth]{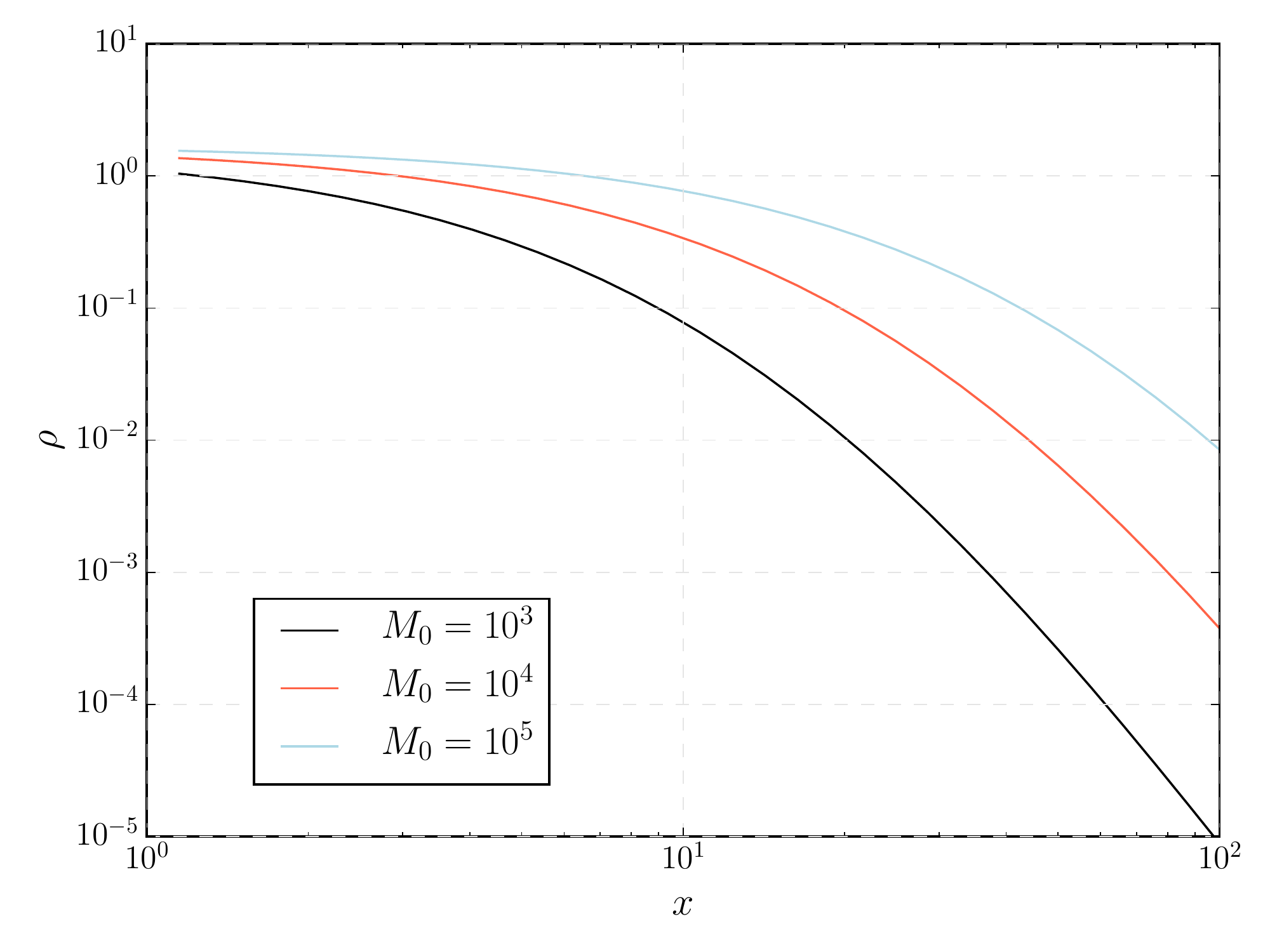}\\
  \includegraphics[width = 0.49\textwidth]{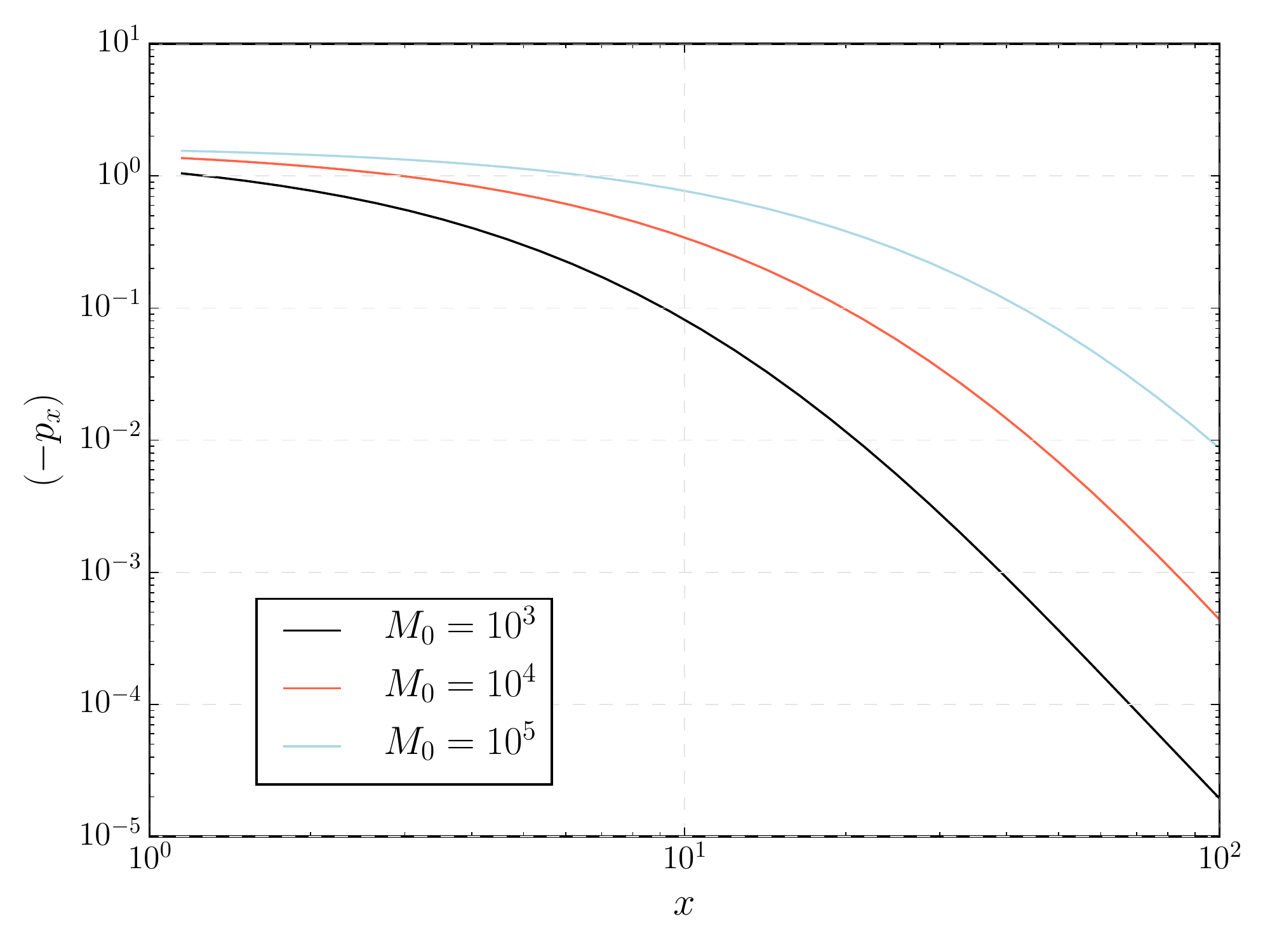}
  \includegraphics[width = 0.49\textwidth]{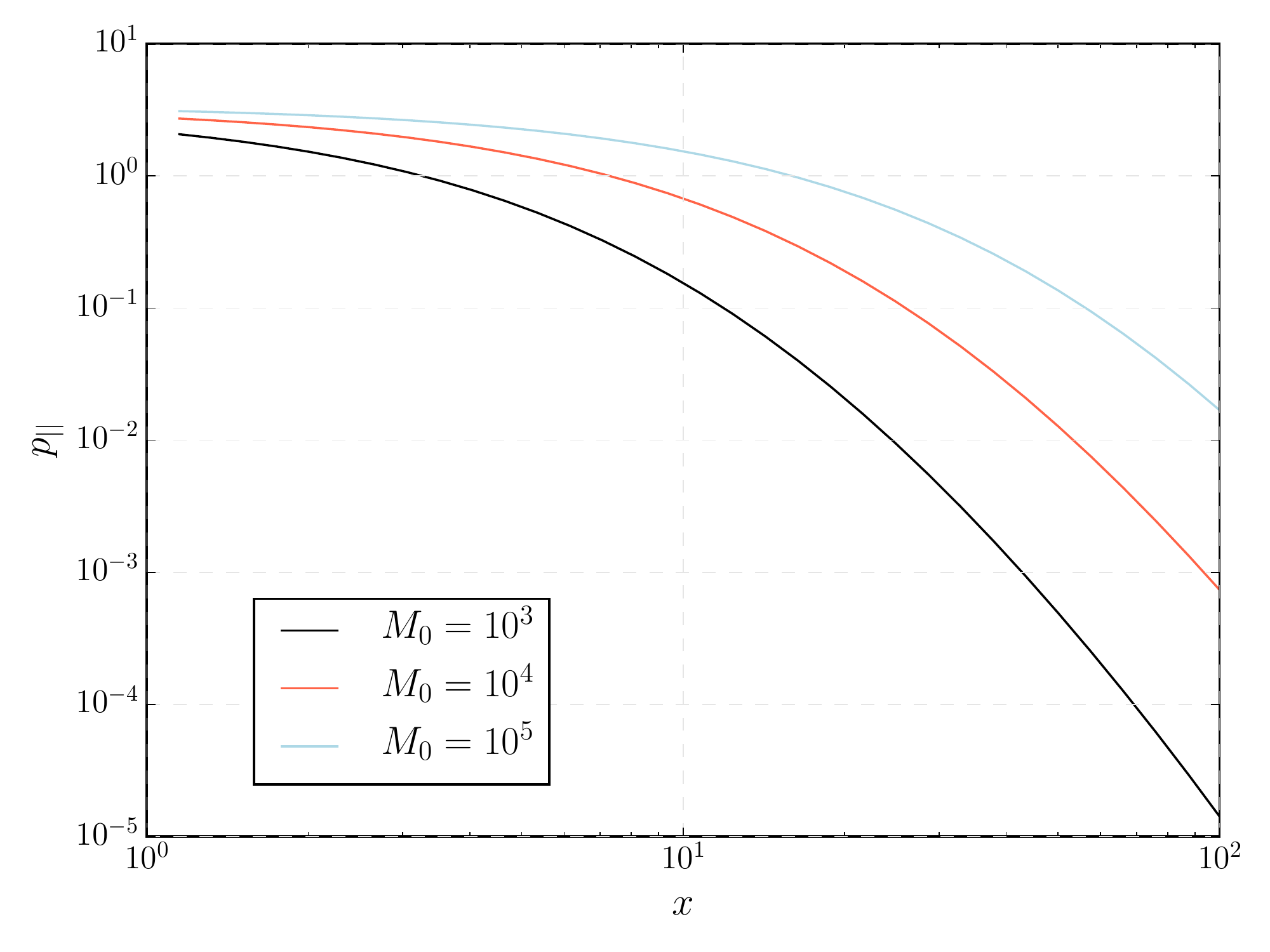}  
}
\caption{Components of the stress-energy tensor for different values of the mass $M_0$.}
\label{fig:tmunu}
\end{figure}

In Fig. \ref{fig:tmunu} we show the behavior of these quantities (as a function of the radial coordinate) in the most quantum region, for different values of the mass. As we see, they reach upper bounds that are order Planck and become mass independent for large $M_0$ (universal upper bounds). The behavior explains the observed properties in the curvature scalars associated with the Ricci tensor, which also show mass-independent (universal) upper bounds.   Therefore, for the quantum states under consideration, these quantities capture the deviations from classical GR. 

\begin{figure}[h]
{\centering     
  \includegraphics[width = 0.79\textwidth]{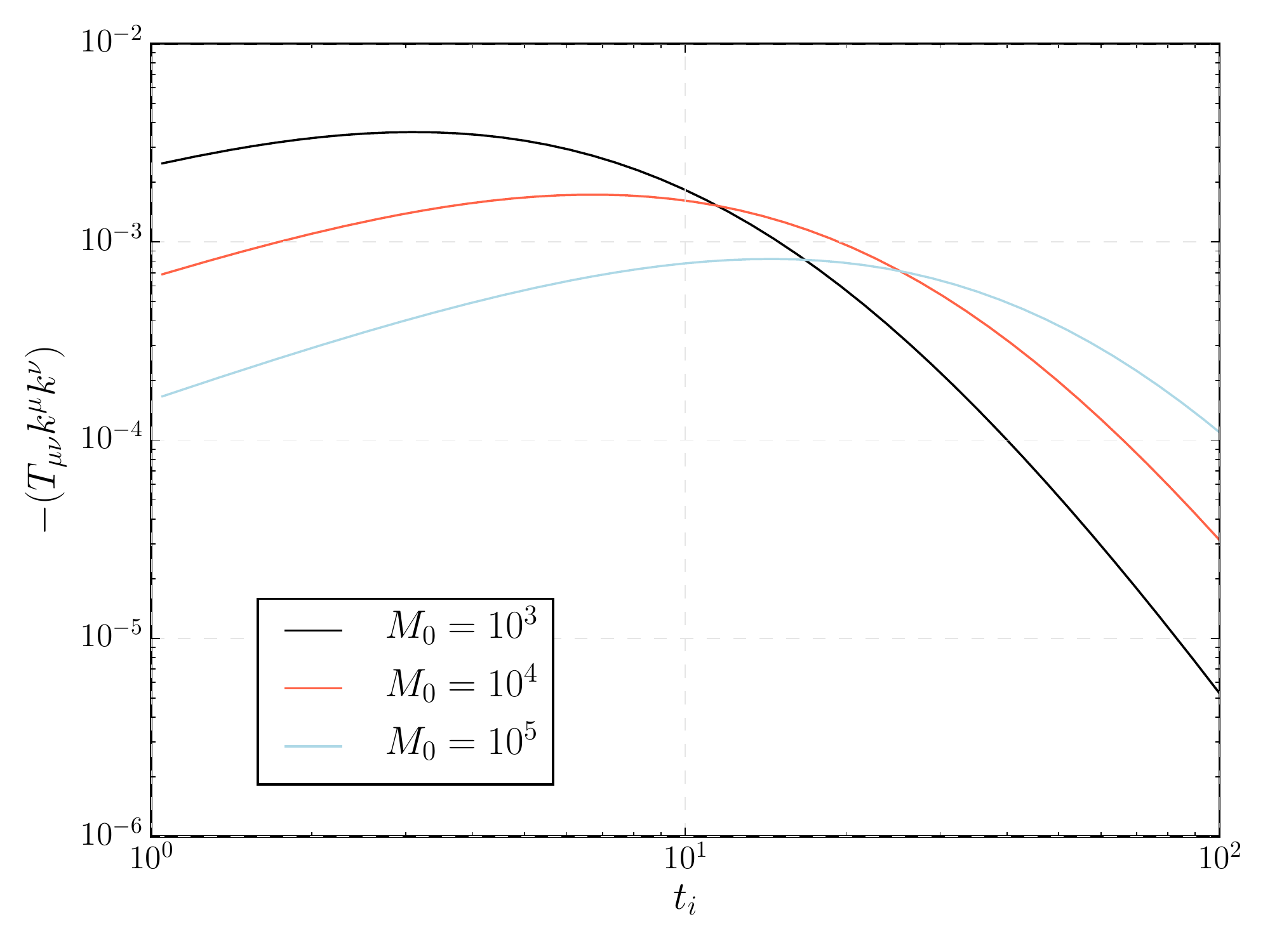} 
}
\caption{Null energy condition for different values of the mass $M_0$ and $\delta x = 1$.}
\label{fig:ene-conds}
\end{figure}

However, since the energy density in the interior is positive, it is not guaranteed that the strong or weak energy conditions are violated. Nevertheless, the energy condition in which the singularity theorems by Hawking and Penrose are based is the null energy condition. We have found null vectors for which the null energy condition in the interior region is violated. Let us, for simplicity, consider a change of coordinates in which the metric in the interior and exterior regions is diagonal. For the exterior region, let us introduce
\begin{equation}
dt_{e} = dt+\frac{{}^{(0)}g_{tx}}{{}^{(0)}g_{tt}}dx.
\end{equation}
On these coordinates, the effective metric takes the simple diagonal form
\begin{equation} \label{eq:gext}
 {}^{(0)}\tilde d\tilde s^2_e = {}^{(0)}\tilde g_{t_e t_e}dt_e^2+{}^{(0)}\tilde g_{xx}dx^2+(x+x_0)^2 d\omega^2,
 \end{equation}
 which makes explicit that ${}^{(0)}\tilde g_{t_e x} = 0$, and where
\begin{align}\nonumber
{}^{(0)}\tilde g_{t_e t_e}&={}^{(0)}\tilde g_{t t}=-\left(1-\frac{r_S}{x+x_0}+\frac{\Delta}{4\pi}\frac{r_S^4 }{(x+x_0)^6\left(1+\frac{r_S}{x+x_0}\right)^2}\right),\\
{}^{(0)}\tilde g_{xx}&={}^{(0)}g_{xx}-\frac{   {}^{(0)}g_{tx}^2 }{ {}^{(0)}g_{tt} }=\frac{\left(1+\frac{\delta x}{2(x+x_0)}\right)^2}{\left(1-\frac{r_S}{x+x_0}+\frac{\Delta}{4\pi}\frac{r_S^4 }{(x+x_0)^6\left(1+\frac{r_S}{x+x_0}\right)^2}\right)}\, ,\label{eq:tildegmunu}
\end{align}
In the interior region, the hypersurfaces $x={\rm const}$ become space-like and homogeneous (we should remember that in the exterior region they are time-like). Then, the coordinate $t_e$ becomes a space-like coordinate, that we will denote by $x_i$, while $x$ is time-like and will be denoted by $t_i$. Then, the metric takes the form 
\begin{equation} \label{eq:gint}
 {}^{(0)}\tilde d\tilde s^2_i = {}^{(0)}\tilde g_{t_i t_i}dt_i^2+{}^{(0)}\tilde g_{x_ix_i}dx_i^2+(t_i+x_0)^2 d\omega^2,
 \end{equation}
with
\begin{align}\nonumber
{}^{(0)}\tilde g_{t_it_i}&=\frac{\left(1+\frac{\delta x}{2(t_i+x_0)}\right)^2}{\left(1-\frac{r_S}{t_i+x_0}+\frac{\Delta}{4\pi}\frac{r_S^4 }{(t_i+x_0)^6\left(1+\frac{r_S}{t_i+x_0}\right)^2}\right)}\, ,\\
{}^{(0)}\tilde g_{x_i x_i}&=-\left(1-\frac{r_S}{t_i+x_0}+\frac{\Delta}{4\pi}\frac{r_S^4 }{(t_i+x_0)^6\left(1+\frac{r_S}{t_i+x_0}\right)^2}\right)\, ,\label{eq:tildegmunu-int}
\end{align}
Now, it is straightforward to construct the null vector
\begin{equation}
k^\mu = \frac{1}{\sqrt{-2{}^{(0)}\tilde g_{t_i t_i}{}^{(0)}\tilde g_{x_i x_i}}}\left(\sqrt{-{}^{(0)}\tilde g_{t_it_i}}\,X^\mu + \sqrt{{}^{(0)}\tilde g_{x_i x_i}}\,r^\mu\right),
\end{equation}
with $X^\mu$ the Killing vector field that is space-like in the interior and $r^\mu$ the time-like vector field normal to the space-like hypersurfaces. Then, we compute $T_{\mu\nu} k^\mu k^\nu$. In Fig. \ref{fig:ene-conds} we see that for this null vector, the corresponding null energy condition is violated. We also see that the violation of this condition weakens with the mass $M_0$. Actually, in the most quantum region ($t_i \ll x_0$) and for $M_0\gg m_{\rm Pl}$, the violation of the null energy condition reaches a maximum that weakens with the mass approximately as $M_0^{-1/3}$.

The emergence of violations of null energy conditions is in line with the elimination of the singularity by loop quantum gravity and gives a hint of how negative energy effects could play a role in other situations where singularities could be eliminated, for instance in wormholes. 
 
\subsection{ADM energy and quasi-local mass expressions}

\subsubsection{ADM energy}

Let us now compute the ADM energy of the effective metric ${}^{(0)}g_{\mu\nu}$. It refers to an asymptotic time translation symmetry related to the time-like Killing vector field in the exterior region. It is defined as follows. From Eqs. \eqref{eq:gext} and \eqref{eq:tildegmunu}, let us introduce the spatial metric ${}^{(0)}q_{ab}$ (where Latin labels refer to spatial indexes):
\begin{equation} \label{q}
 {}^{(0)}q_{ab} d x^{a} d x^{b} =\frac{\left(1+\frac{\delta x}{2(x+x_0)}\right)^2}{\left(1-\frac{r_S}{x+x_0}+\frac{\Delta}{4\pi}\frac{r_S^4 }{(x+x_0)^6\left(1+\frac{r_S}{x+x_0}\right)^2}\right)}dx^2 + (x+x_0)^2 d \omega^{2}  \, ,
\end{equation}
Besides, the lapse ${}^{(0)}N$ is defined as 
\begin{equation}
{}^{(0)}N=\sqrt{-{}^{(0)}g_{t_et_e}} = \sqrt{1-\frac{r_S}{x+x_0}+\frac{\Delta}{4\pi}\frac{r_S^4 }{(x+x_0)^6\left(1+\frac{r_S}{x+x_0}\right)^2}}.
\end{equation}
The ADM energy is then given by (see, e.g., \cite{tt}):
\begin{equation}
E_{\rm ADM}=\lim_{x\to\infty}\frac{1}{16 \pi G} \oint_{x} dS_{d}\, \big(\det {}^{(0)}q \big)^{\frac{1}{2}} \,{}^{(0)}q^{ac} {}^{(0)}q^{bd} \,\big[ {}^{(0)}N \partial_ {[c} {}^{(0)}q _ {b]a} - \left({}^{(0)}q_{a[b} - {\delta}_{a[b}\right) ({\partial}_{c]} {}^{(0)}N) \big]\, ,  
\end{equation}
where partial derivatives refer to the spatial Cartesian coordinates of the obvious Minkowski ${}^{(0)}\eta^{o}_{\mu\nu}$ associated with ${}^{(0)}g_{\mu\nu}$. Substituting for  ${}^{(0)}q_{ab}$ from (\ref{q}) we obtain
\begin{equation}
\lim_{{x} \to\infty}\frac{1}{16 \pi G} \oint_{x} dS_{d}\, \big(\det {}^{(0)}q\big)^{\frac{1}{2}} \,{}^{(0)}q^{ac} {}^{(0)}q^{bd} \,\big[ {}^{(0)}N \partial_ {[c} {}^{(0)}q_{b]a}\big]
= M_0+\frac{\delta x}{2G},
\end{equation}
and 
\begin{equation}
\lim_{{x} \to\infty}\frac{1}{16 \pi G} \oint_{x} dS_{d}\, \big(\det {}^{(0)}q\big)^{\frac{1}{2}} \,{}^{(0)}q^{ac} {}^{(0)}q^{bd} \,\big[\left({}^{(0)}q_{a[b} - {\delta}_{a[b}\right) ({\partial}_{c]} {}^{(0)}N) \big] =0\, ,
\end{equation}
Therefore, the ADM mass agrees very well with $M_0$ except for a small correction whose relative value is $\delta x/(2GM_0)$. For macroscopic black holes, this correction is negligible (tends to zero in the limit $M_0\to \infty$). We have also checked that expressions for the ADM mass in terms of the three dimensional Ricci tensor (see for instance Eq. (3.4) of Ref. \cite{am}) give the same result, as one should expect from the fall-off properties of our effective metric. 

\subsubsection{Komar mass}

We have also computed the Komar mass. The general expression is given by (see Ref. \cite{wald,jg})
\begin{equation}
M_K=-\frac{1}{8\pi}\int_{{\cal S}^2}\epsilon^{\mu\nu\rho\gamma}\nabla_\mu X_\nu({\rm d}\Omega)_{\rho\gamma},
\end{equation}
where $X^\mu$ is the Killing vector field that is time like in the exterior region, $\nabla_\mu$ the connection compatible with our effective metric, $\epsilon_{\mu\nu\rho\gamma}$ the total antisymmetric tensor and $({\rm d}\Omega)_{\mu\nu}$ is the (2-form) surface element of the 2-sphere ${\cal S}^2$ where the integral is computed. The Komar mass, for our effective metric in the $(t_e,x,\theta,\phi)$ coordinates, takes a simple form 
\begin{equation}
M_{K} = M_0\frac{1}{\left(1+\frac{\delta x}{2(x+x_0)}\right)}\left(1-\frac{\Delta}{2\pi}\frac{8G^3M_0^3}{(x+x_0)^5}\frac{3+\frac{4 GM_0}{(x+x_0)}}{\left(1+\frac{2GM_0}{x+x_0}\right)^3}\right).
\end{equation}
As we see, in the limit $\delta x\to 0$ and $\Delta\to 0$, the Komar mass reduces to $M_0$. This agreement is also achieved in the limits $x\gg x_0$. Thus, we can interpret $M_0$ as the Komar mass at spatial infinity. Let us notice that the ADM mass and the Komar mass agree up to a correction proportional to $\delta x/(2G M_0)$ at spatial infinity, which, in relative terms, is negligible for macroscopic black holes. 

\begin{figure}[h]
{\centering     
  \includegraphics[width = 0.79\textwidth]{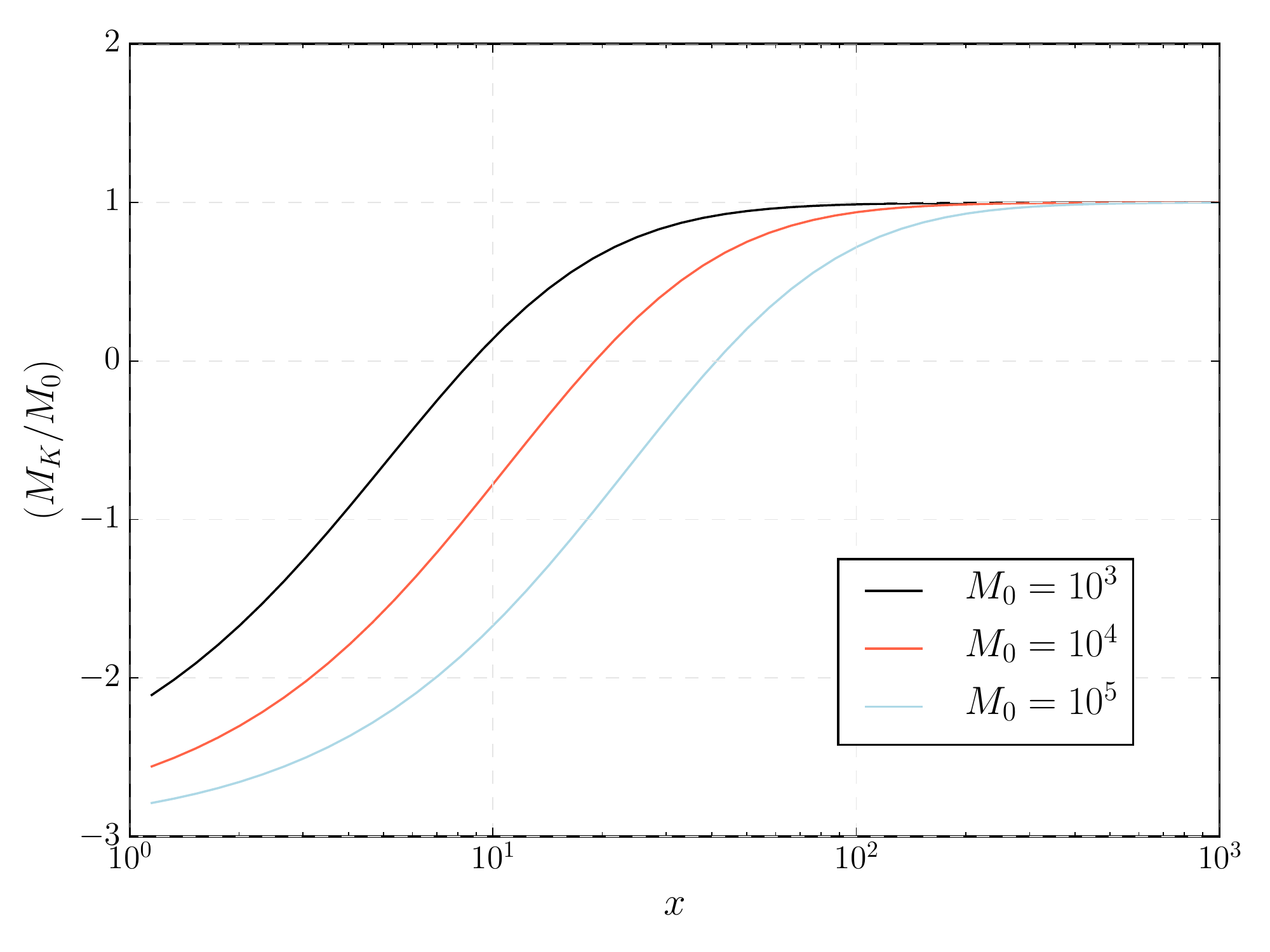}
}
\caption{Komar mass (normalized to the corresponding mass $M_0$) for different values of $M_0$. }
\label{fig:komar}
\end{figure}

On the other hand, in the limit $x\ll x_0$ and $M\gg m_{\rm Pl}$, we obtain
\begin{equation}\label{eq:km-limit}
M_{K} = -3M_0\left(1+{\cal O}(M_0^{-1/3})\right).
\end{equation}
This negative lower bound in the Komar mass can be explained by the positivity of the effective energy density and the fact that the Komar mass approaches $M_0$ in the low curvature region. Therefore, the Komar mass must be negative in the most quantum region in order to compensate the positive contribution of $\rho$ accumulated as one moves towards spatial infinity. In Fig. \ref{fig:komar} we show the Komar mass (normalized to $M_0$) for several choices of $M_0$. As we mentioned, the Komar mass is negative when we approach the most quantum region, i.e. $x\leq x_0$, in agreement with the limit \eqref{eq:km-limit}. 

\subsubsection{Misner--Sharpe mass}

Another interesting definition of quasi-local mass for spherically symmetric spacetimes is the so-called Misner--Sharpe mass \cite{ms,jg}. It is defined as 
\begin{equation}
M_{MS}=-\frac{r^3}{8G}R_{\mu\nu\rho\gamma}\epsilon^{\mu\nu}\epsilon^{\rho\gamma}
\end{equation}
where $\epsilon_{\mu\nu}=\hat X^\rho \hat r^\gamma\epsilon_{\rho\gamma\mu\nu}$ with $\hat X^\mu $ and $\hat r^\mu$ the normalized vectors normal to the 2-spheres of constant radius $r$. The explicit form in the $(t_e,x,\theta,\phi)$ chart is
\begin{equation}
M_{MS}=-\frac{x+x_0}{2G\left(1+\frac{\delta x}{2(x+x_0)}\right)^2}\left(1-\frac{1-\frac{r_S}{x+x_0}+\frac{\Delta}{4\pi}\frac{r_S^4 }{(x+x_0)^6\left(1+\frac{r_S}{x+x_0}\right)^2}}{\left(1+\frac{\delta x}{2(x+x_0)}\right)^2}\right)
\end{equation}
It is straightforward to check that in the limit $\delta x\to 0$ and $\Delta\to 0$, the Misner--Sharpe mass reduces to $M_0$. Therefore, the ADM, Komar and Misner--Sharpe masses agree in the classical limit. When quantum corrections are present, in the limit $x\to \infty$ the Misner--Sharpe mass equals 
\begin{equation}
\lim_{x\to\infty}M_{MS}=M_0+\frac{\delta x}{2G}.
\end{equation}
Hence, the Misner--Sharpe and the ADM mass agree at spatial infinity. 

\begin{figure}[h]
{\centering     
  \includegraphics[width = 0.79\textwidth]{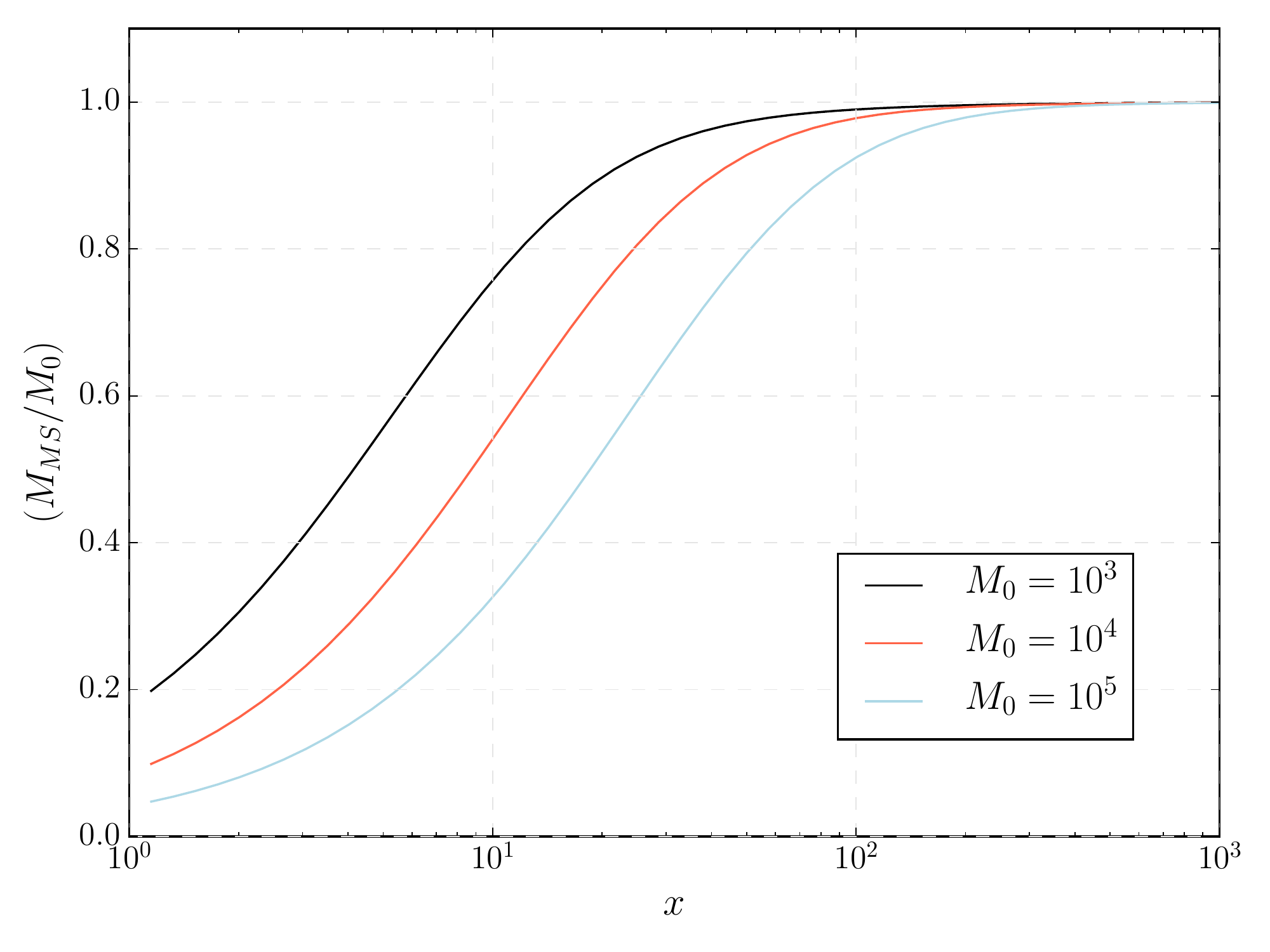}
}
\caption{Misner--Sharpe mass (normalized to the corresponding mass $M_0$) for different values of $M_0$. }
\label{fig:misner-sharpe}
\end{figure}

On the other hand, in the most quantum region, for $x\ll x_0$ and in the limit $M_0\gg m_{\rm Pl}$, 
\begin{equation}\label{eq:ms-limit}
M_{MS} = \frac{1}{2G}\left(\frac{4 G M_0 \Delta }{\pi}\right)^{1/3}\left(1+{\cal O}(M_0^{-1/3})\right).
\end{equation}
In Fig. \ref{fig:misner-sharpe} we show the Misner--Sharpe mass (normalized with respect to $M_0$) for several choices of $M_0$. As we mentioned, this mass grows as $M_0^{1/3}$ in the most quantum region. Therefore, the ratio $M_{MS}/M_0$ tends to zero there in the limit $M_0\gg m_{\rm Pl}$, in agreement with what is shown in Fig. \ref{fig:misner-sharpe}. 

We conclude that these different notions of (quasi-local) energy, although they give results that agree in the classical limit, this is not the case when quantum corrections are present.

\section{Discussion}\label{sec:concl}

In this manuscript we have studied the loop quantization of spherically symmetric space-times. Our  treatment adopts the improved dynamics scheme of Chiou {\em et al.} \cite{chiou}, similar to the ones typically adopted in LQC. We identify suitable physical operators for the space-time metric components along with a suitable family of semiclassical physical states, derive the corresponding effective geometries, and study their properties. We found i) the classical singularity is replaced by a regular but discrete region where curvature is high, ii) in the most quantum region curvatures reach universal upper bounds that are at most Planck-scale and not trans-Planckian as in previous quantizations, and iii) at low curvatures the effective space-time approaches the Schwarzschild geometry sufficiently fast. This is captured in part in the agreement that we found between several asymptotic and quasi-local notions of mass evaluated at spatial infinity. All of them give the same finite results (except for very tiny quantum corrections). 

These effective geometries share several properties with the recent proposal of Ref. \cite{aos}. The most obvious one is the universal upper bounds in the curvature in the most quantum region. However, there are also a number of differences. Our effective geometries have not been extended yet beyond the high curvature regions (although extensions are in principle viable and will be studied in a future publication). They encompass {\em simultaneously} one of the exterior regions and the trapped (interior) region all the way until curvature reaches a maximum magnitude. The stress-energy tensor of our model also violates energy conditions, but in a weaker sense with respect to the corresponding one in Ref. \cite{aos}. It also falls off sufficiently fast at low curvatures. This considerably affects the behavior of the Komar mass in the high and low curvature regions. Actually, our effective geometries approach the Schwarzschild metric sufficiently fast at spatial infinity (see Ref. \cite{ao} for a recent discussion about the asymptotic behavior of \cite{aos}). 

The effective geometries studied in this manuscript have been derived from the quantum theory for a suitable family of semiclassical states. However, more general states can be explored within this quantization. We expect that additional phenomenological aspects, like fluctuations of the mass and graphs, will contribute to the effective geometries. Moreover, it will be interesting in the future to explore slicings that allow us to extend our effective geometries beyond the most quantum region, and construct the corresponding full Penrose diagram.  Besides, these ideas can be easily extended to other scenarios like Reissner-Nordstr\"om black holes \cite{us-rn} or even dynamical scenarios of black hole formation \cite{shell}. All these aspects will be a matter of future research. 

\section*{Acknowledgment}
 This work was supported in part by Grants NSF-PHY-1603630,
NSF-PHY-1903799, funds of the Hearne Institute for Theoretical
Physics, CCT-LSU, fqxi.org, Pedeciba and Fondo Clemente Estable
FCE\_1\_2019\_1\_155865, and Project. No. FIS2017-86497-C2-2-P of MICINN from Spain. J.O. acknowledges the Operative Program FEDER 2014-2020 and the Consejer\'ia de Econom\'ia y Conocimiento of the Junta de Andaluc\'ia.

\appendix

\section{Eddington--Finkelstein (E-F) coordinates in classical GR}\label{app:class-EF}

In order to describe spherically symmetric geometries in a horizon penetrating slicing, we consider Eddington--Finkelstein (E-F) coordinates. They cover half of the Kruskal diagram, including the exterior and the interior regions of either the (future) black or (past) white hole.

 In the canonical framework this amounts to the following gauge fixing condition,
 \begin{equation}
 \Phi_1^{\eta}=K_\varphi-\eta\frac{r_S}{\sqrt{E^x(x)}}\frac{1}{\sqrt{1+\frac{r_S}{\sqrt{E^x(x)}}}},\quad \eta=\pm1.
 \end{equation}
It corresponds to either outgoing ($\eta=+1$) or ingoing ($\eta=-1$) E-F coordinates. The dynamical preservation of this condition allows us to fix the lapse function. After solving both $\Phi_1^{\eta}=0$ for $K_\varphi$ and the scalar constraint as
 \begin{equation}\label{eq:ephi}
[E^\varphi]^2=\frac{([E^x]')^2}{4}\left(1+\frac{r_S}{\sqrt{E^x(x)}}\right),
 \end{equation}
 the lapse function (squared) takes the form
 \begin{equation}
N^2=\frac{1}{1+\frac{r_S}{\sqrt{E^x(x)}}}.
 \end{equation}
 After the gauge fixing, the reduced Hamiltonian is
\begin{equation}\label{eq:total-ham2}
\tilde H_T=\int dx \tilde N^x\tilde H_x,
\end{equation} 
with $\tilde H_x$ still given by Eq. \eqref{eq:difeo} but substituting $E^\varphi$ by \eqref{eq:ephi} and $K_\varphi$ after solving $\Phi_1^{\eta}=0$. We now consider the gauge fixing condition
 \begin{equation}
\Phi_2=E^x(x)-g(x),
 \end{equation}
 where $g(x)$ is an arbitrary function but with $g'(x)\neq 0$ for all $x$ (typically in E-F coordinates one chooses $g(x)=x^2$). Preservation of this gauge fixing, i.e. $\dot \Phi_2 = 0$, yields
 \begin{equation}
N^x=-\frac{2\,\eta \,r_S\,g(x)}{g'(x)}\frac{1}{1+\frac{r_S}{\sqrt{g(x)}}},
 \end{equation}
with $E^x(x)=g(x)$ replaced everywhere and $K_x$ (easily) determined by the condition $\tilde H_x=0$. 

\section{Superpositions in the mass}\label{app2}

In order to understand the effects of fluctuations of the mass on the effective geometries, let us consider the states
\begin{equation}\label{eq:gauss}
\psi(M)=\frac{1}{\Delta M_0}e^{iM P_0/\hbar}\cos\left[\frac{\pi(M-M_0)}{2\Delta M_0}\right]\Theta(M-M_0+\Delta M_0)\Theta(M_0+\Delta M_0-M).
\end{equation}
One can see that
\begin{equation}
\langle \hat M\rangle = M_0,\quad \langle \hat M^2 \rangle = M_0^2+\Delta M_0^2\left( \frac{1}{3} - \frac{2 }{\pi^{2}}\right).
\end{equation}
If we want to compute integrals of the form 
\begin{equation}
\int dM |\psi(M)|^2\,F(M) 
\end{equation}
for some well-defined function $F(M)$ around $M\in[M_0-\Delta M_0,M_0+\Delta M_0,]$, it is convenient to introduce $y=(M-M_0)/\Delta M_0$ and Taylor expand $F(y \Delta M_0+M_0)$ for $\Delta M_0$ small. Concretely,
\begin{equation}
F(M)=\sum_{n=0}^\infty \frac{1}{n!}\frac{d^nF(M_0)}{dM_0^n} \Delta M_0^n y^n.
\end{equation}
Then, the integrals reduce to
\begin{align}
  &\int dM |\psi(M)|^2\,F(M) = \int \frac{dy}{\Delta M_0} |\psi(\Delta M_0\,y+M_0)|^2 F(\Delta M_0 \,y + M_0) \\
&= \sum_{n=0}^\infty \frac{1}{n!}\frac{d^nF(M_0)}{dM_0^n} \Delta M_0^n\int_{-1}^1 dy \frac{1}{2}\left[1+\cos\left(\pi \,y\right)\right] y^n.
\end{align}
These integrals are known for any finite $n$. However, here we are interested in the leading and subleading contributions in $\Delta M_0$. In total, we get
\begin{equation}
\int dM |\psi(M)|^2\,F(M) = F(M_0)+\frac{1}{2}F''(M_0) \Delta M_0^2\left(\frac{1}{3}-\frac{2}{\pi^2}\right)+{\cal O}(\Delta M_0^4).
\end{equation}
where the primes denote derivatives with respect to the argument $M$ of the function $F(M)$. 

\begin{figure}[h]
{\centering     
  \includegraphics[width = 0.49\textwidth]{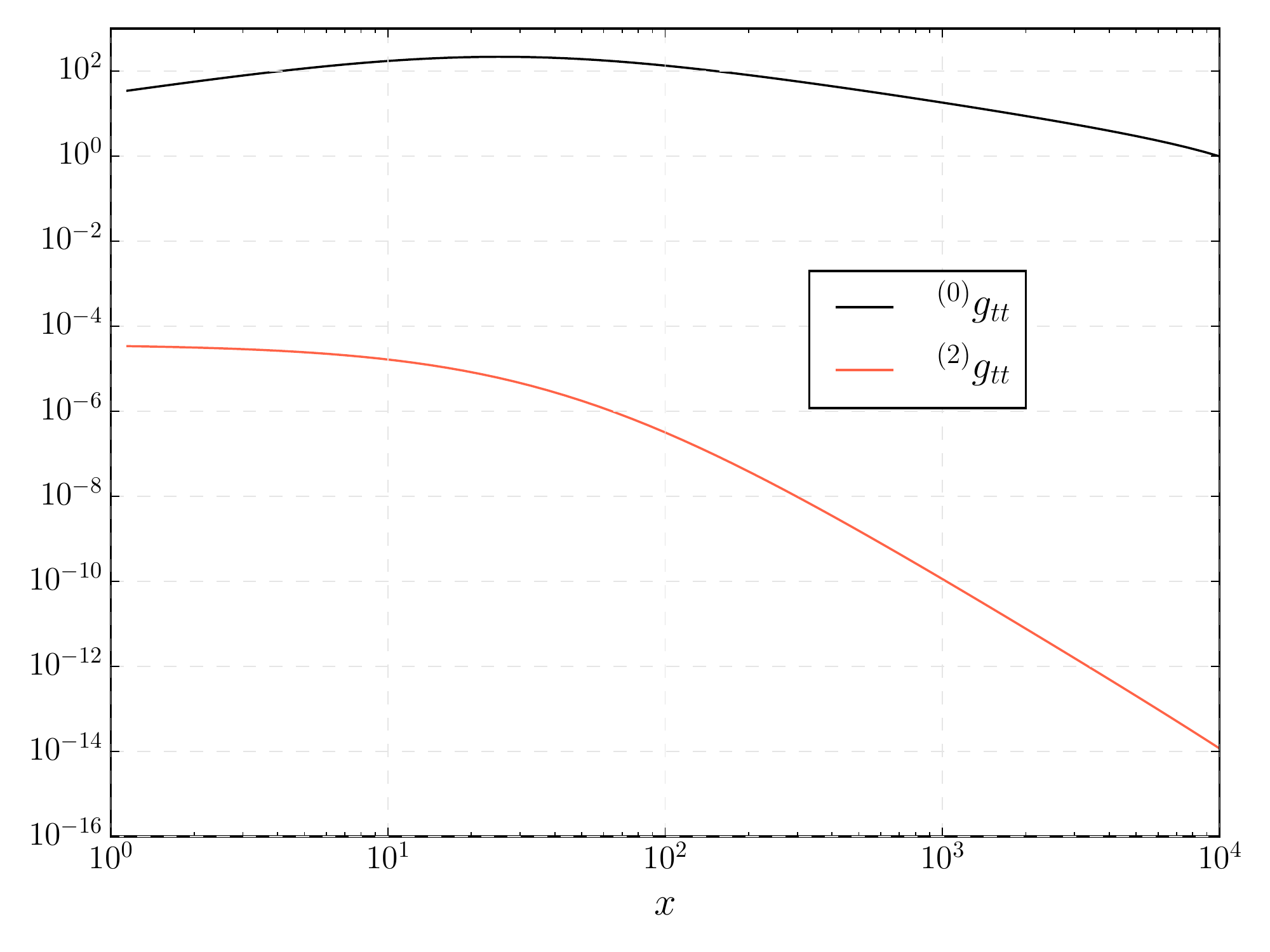} 
  \includegraphics[width = 0.49\textwidth]{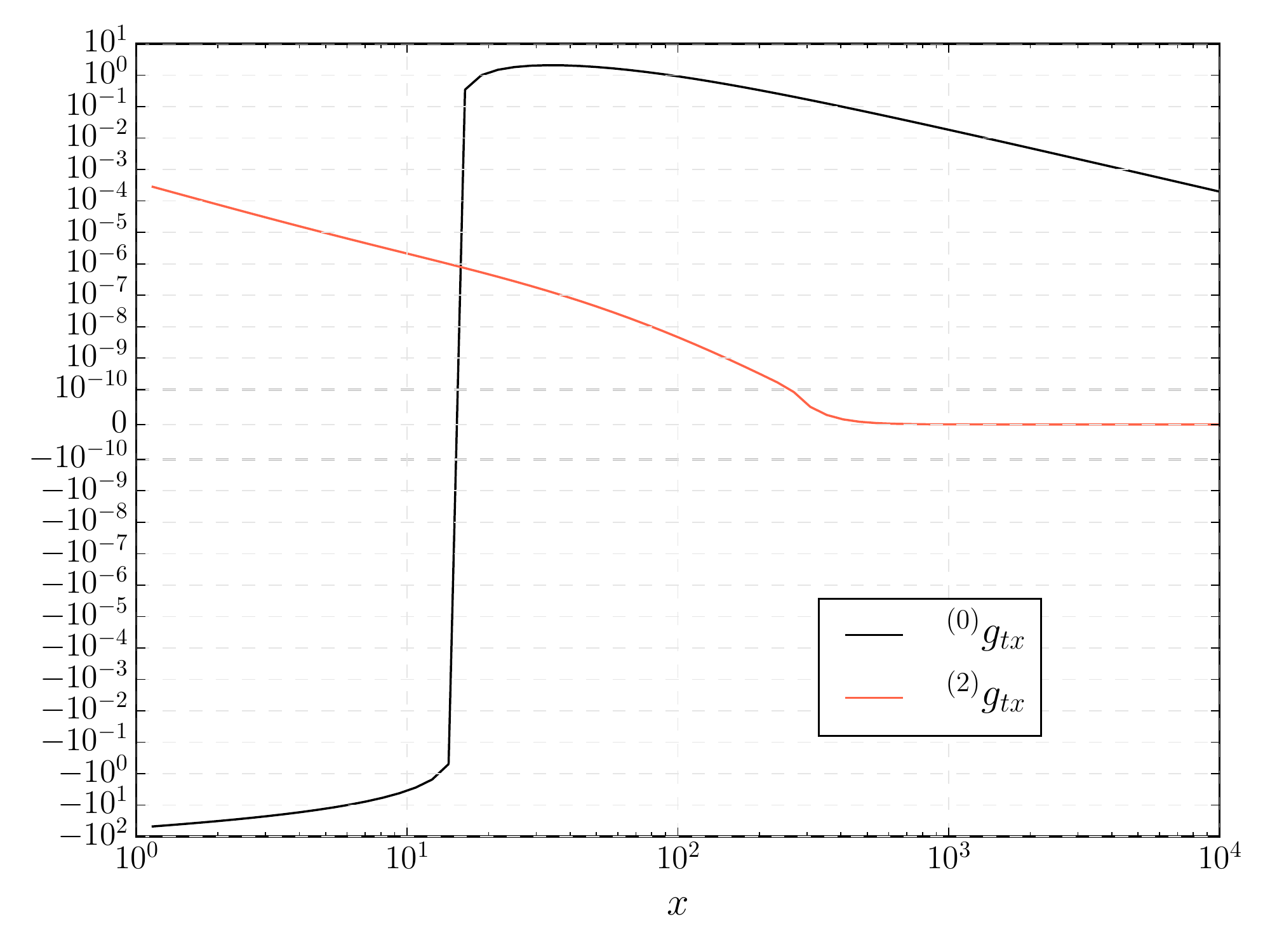}
}
\caption{Components of the effective metric without fluctuations of the mass and the next to leading order corrections in presence of these fluctuations. Here we consider $M_0=10^4$ and $\Delta M_0=\widetilde{\Delta M_0}=7.9$, both in Planck units.}
\label{fig:gmunu}
\end{figure}

We will adopt this expansion for the expectation values of the operators representing components of the space-time metric defined by \eqref{eq:hatgmunu}. Let us recall that the effective metric can be expressed as $g_{\mu\nu}={}^{(0)}g_{\mu\nu}+{}^{(2)}g_{\mu\nu}\Delta M_0^2+\ldots$, where the leading order contribution turns out to be given by Eq. \eqref{eq:g0}, while the subleading contribution ${}^{(2)}g_{\mu\nu}$ to the effective metric is given by
\begin{equation}
{}^{(2)}g_{tt} = \left(\frac{1}{3}-\frac{2}{\pi^2}\right)\frac{d^2}{dr_S^2}{}^{(0)}g_{tt}\bigg|_{r_S=2GM_0},\quad {}^{(2)}g_{tx} = \left(\frac{1}{3}-\frac{2}{\pi^2}\right)\frac{d^2}{dr_S^2}{}^{(0)}g_{tx}\bigg|_{r_S=2GM_0}.\label{eq:hatgmunu2}
\end{equation}
The other contributions ${}^{(2)}g_{xx}=0$ as well as ${}^{(2)}g_{\theta\theta}=0={}^{(2)}g_{\phi\phi}$. 
In Fig. \ref{fig:gmunu} we show the components of the effective metric ${}^{(0)}g_{\mu\nu}$ studied in the main body of this manuscript, together with the next to leading order contribution  ${}^{(2)}g_{\mu\nu}$, for $M_0=10^4$ and $\Delta M_0=\widetilde{\Delta M_0}$ where, again, 
\begin{equation}
\widetilde{\Delta M_0} = \frac{3}{2}\left(\frac{4\pi \ell_{\rm Pl}^3}{2G\Delta}\right)^{2/3}M_0^{1/3},
\end{equation}
is the largest dispersion in the mass compatible with the family of semiclassical states considered in our manuscript. 
These results confirm that contributions to the effective metric due to $\Delta M_0$ are small for the particular family of states under consideration. However, less semiclassical states with larger fluctuations in the mass could modify considerably the effective metric. 

\section{Continuum limit}\label{app:cont}

In this appendix we compare the continuous and discrete second derivatives of the effective metric components. For this purpose, we compare $\partial_x^2 {}^{(0)}g_{\mu\nu}$ with its discrete version
\begin{equation}
  \Delta_{\delta x}^2 {}^{(0)}g_{\mu\nu}(x_j)=\frac{{}^{(0)}g_{\mu\nu}(x_j)-2{}^{(0)}g_{\mu\nu}(x_j+\delta x)+{}^{(0)}g_{\mu\nu}(x_j+2\delta x)}{\delta x^2}.
\end{equation}

\begin{figure}[h]
{\centering     
  \includegraphics[width = 0.49\textwidth]{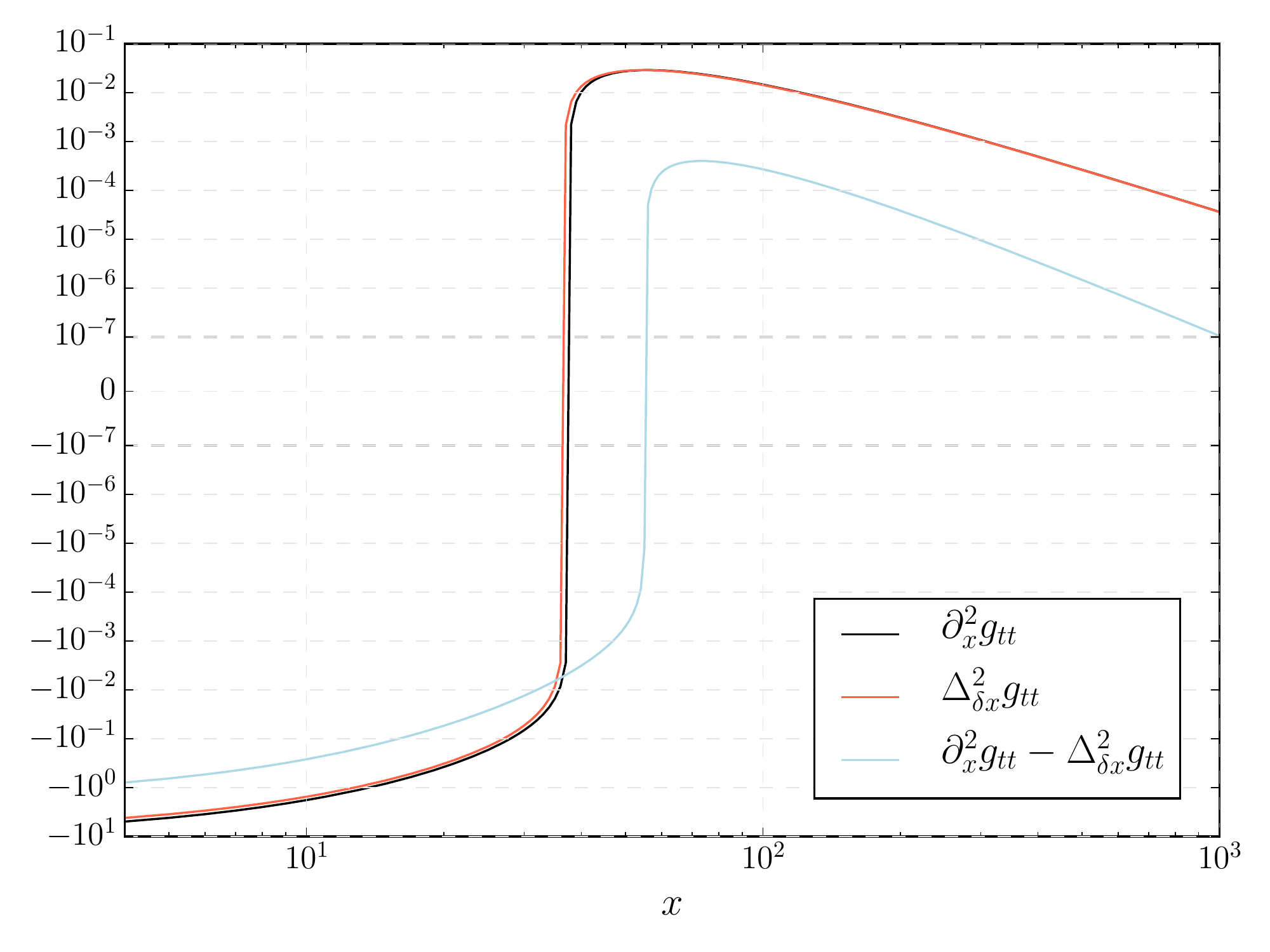}\\ 
  \includegraphics[width = 0.49\textwidth]{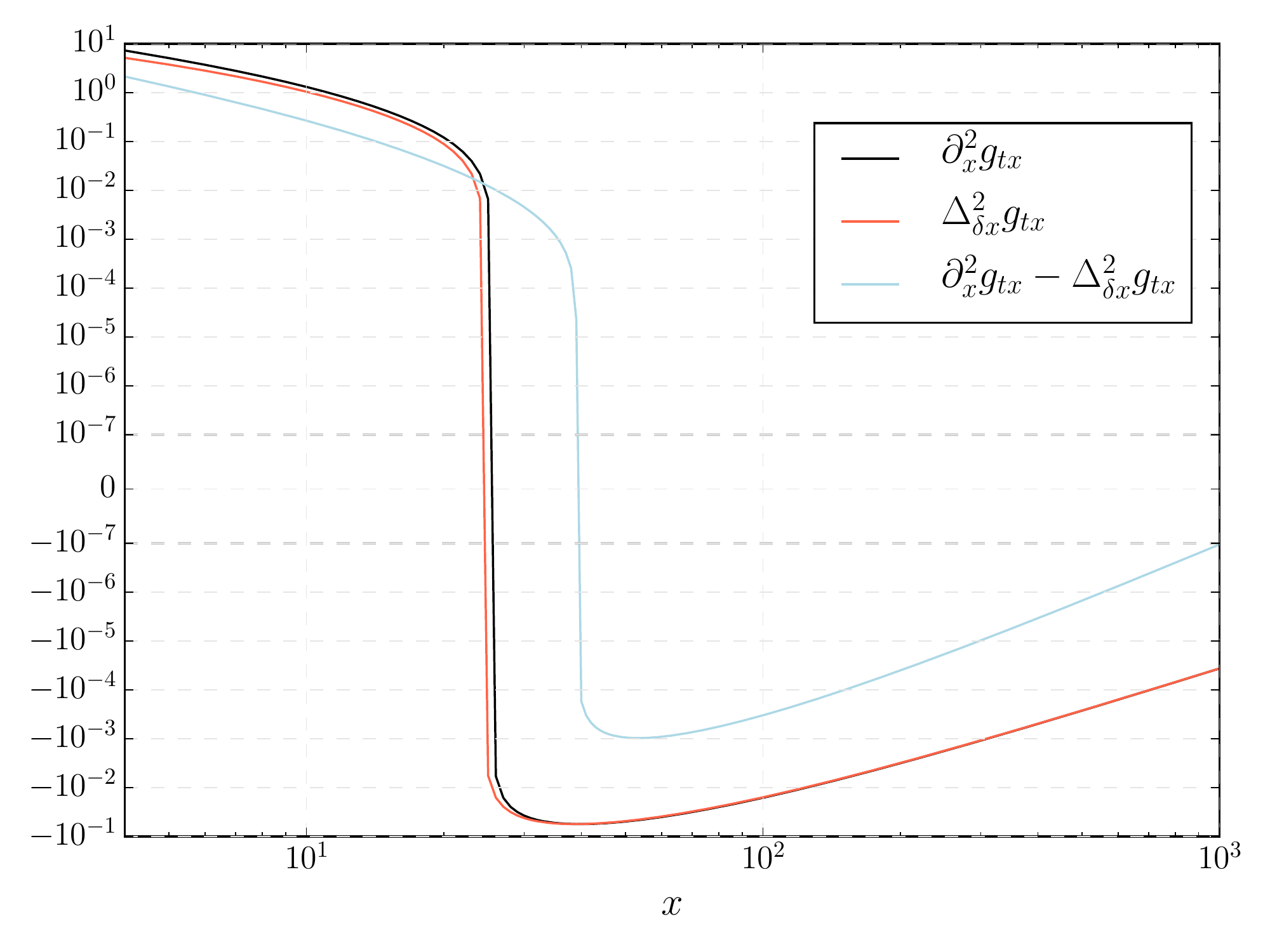}
  \includegraphics[width = 0.49\textwidth]{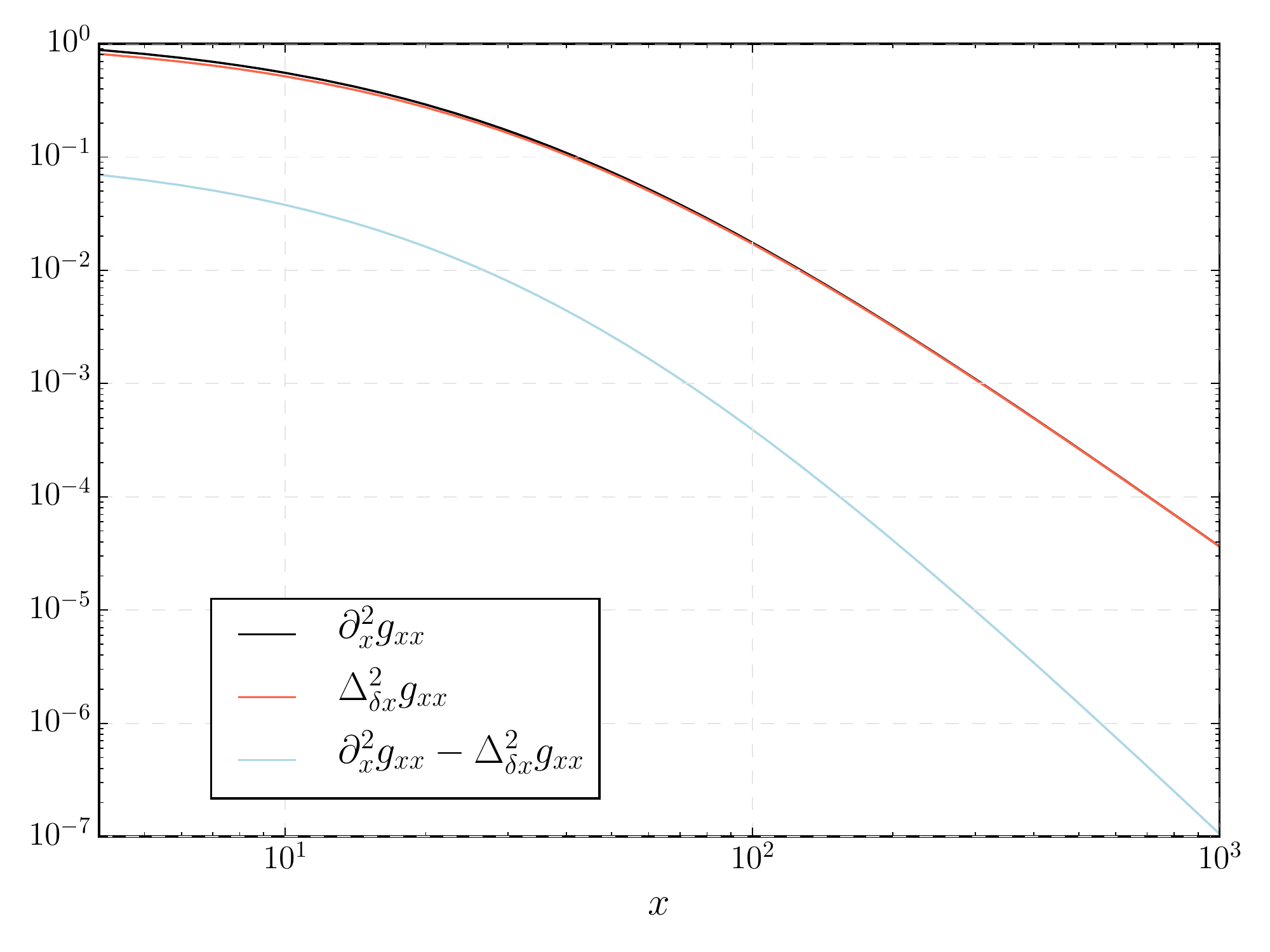}  
}
\caption{Comparison of the second order discrete and continuous spatial derivatives of some of the components of the effective metric. In addition, we choose $M_0=10^4$ and $\delta x=1$ in Planck units.}
\label{fig:ddgmunu}
\end{figure}

It is very easy to verify that this effective metric satisfies $\Delta_{\delta x}^2 {}^{(0)}g_{\theta\theta}(x_j)=2=\partial_x^2{}^{(0)}g_{\theta\theta}(x_j)$. Similarly, $\Delta_{\delta x}^2 {}^{(0)}g_{\phi\phi}(x_j)=2\sin^2\theta=\partial_x^2{}^{(0)}g_{\phi\phi}(x_j)$. In Fig. \ref{fig:ddgmunu}, we compare both second discrete and continuous derivatives of the remaining metric components for $\delta x=\ell_{\rm Pl}$. 

We see that the error we make in approximating discrete derivatives by continuous ones is around $10\%$ in the most quantum region, but this (relative) error decreases as we move to the low curvature region.

\end{document}